\documentclass[11pt]{article}

\usepackage[]{graphics,graphicx}

\usepackage{amsmath}
\usepackage{pstricks}

\usepackage{blindtext}

\newcommand{\Pd}[2]{\frac{\partial #1}{\partial #2}}
\newcommand{\lb}{\left(}
\newcommand{\rb}{\right)}
\newcommand{\D}{\displaystyle}

\usepackage{setspace} 

\RequirePackage{lineno}

\begin{document} 

 \modulolinenumbers[1]

%\linenumbers
%\baselineskip24pt

\begin{center}
{\Large\bf Unified Mechanical Erosion Model for Multi-phase Mass Flows}
\\[10mm]
{Shiva P. Pudasaini
\\[3mm]
Technical University of Munich, 
School of Engineering and Design,\\
Chair of Landslide Research,}\\
{Arcisstrasse 21, D-80333, Munich, Germany}\\[3mm]
Kathmandu Institute of Complex Flows,\\ Kageshwori Manohara - 3,
Bhadrabas, Kathmandu, Nepal\\[1mm]
{E-mail: shiva.pudasaini@tum.de}\\[7mm]
\end{center}
\noindent
\noindent
{\bf Abstract:} 
Erosion often poses a great challenge in simulating hazardous multi-phase mass flows as it can drastically change the flow behavior, impact force, run-out and deposition morphology by dramatically increasing their masses, potentially adversely affecting the population and civil structures. However, there exists no mechanically-explained process-based, unified multi-phase erosion model. Here, based on the comprehensive descriptions of the frictional, collisional and viscous stress generating mechanisms, we construct a novel, unified and consistent mechanical basal erosion rates for solid and fluid phases and demonstrate their richness, urgency and wide spectrum of applicability. This is achieved by seminally introducing mechanically interacting stress structures across the erosion-interface between the landslide and the bed. All the shear resistances from the bed against the applied shear stresses from the landslide are based on consistent physical principles. The proposed multi-phase interactive shear structures are mechanically superior, dynamically flexible and broader over the existing models. As the total erosion rate is the exact sum of the solid and fluid erosion rates; which are extensive, compact and mechanically fully described; this automatically satisfies the natural criterion. These erosion rates consistently take the solid and fluid fractions from the bed which are customarily supplied to the solid and fluid components in the flow. This overcomes severe limitations inherited by existing erosion models, and thus opens ample possibilities for real applications. For the first time, we physically correctly model the essentially composite, intricate erosion velocities of the mobilized particles and fluid from the bed and utilize them to architect erosion-induced net momentum productions that include all the interactions between solids and fluids in the landslide and the bed. With this, we construct complete momentum productions for solid and fluid phases. Both the erosion rates and momentum productions are well constrained. We invent the stress correction factor, erosive-shear-velocity, super-erosion-drift, and the erosion-matrix which inherently characterize complex erosion processes in multi-phase mass flows. This greatly enhances our understanding. By embedding extensive erosion velocities, unified mechanical erosion rates and the advanced net momentum productions into the mass and momentum balance equations, we develop a novel, realistic and mechanically-explained, comprehensive multi-phase model for erosive mass flows. Thus, our approach makes a complete description of multi-phase erosive landslide dynamics by considering all essential aspects associated with erosion including the correct handling of inertia. As the new model covers a broad spectrum of natural processes, it offers great opportunities for practitioners in appropriately solving technical, engineering and geomorphological problems related to complex erosive multi-phase mass flows.

\section{Introduction}

Dynamics of geophysical mass flows including landslides, avalanches and
debris flows can be dominantly affected by the complex mechanical
processes of erosion, entrainment and deposition (Huggel et al., 2005; Hungr et al., 2005; Santi et al., 2008; de Haas et al., 2016, 2020).
Here, the terms landslides, avalanches, debris flows and mass flows are used as synonyms.
As these multi-phase flows cascade down mountain slopes the sediment and fluid are entrained from the bed. In turn, these events can disproportionately increase their volumes and destructive potentials by several orders of magnitude and become exceptionally mobile (Evans et al., 2009; Theule et al., 2015; Somos-Valenzuela et al., 2016; Mergili et al., 2018; Liu et al., 2019).
{Erosion,} entrainment and associated flow bulking in landslide prone areas and debris-flow torrents are a major concern for civil and environmental engineers and landuse planners. {It requires a cost-intensive mitigation of the associated hazard}. 
Mobility is among the most important features of the erosive landslide as it directly measures the threat posed by the landslide. Landslide mobility is associated with erosion-induced excessive volume and material properties and is characterized by {an} enormous impact {force}, exceptional travel distance and inundation area. 
{Thus,} erosion$-$induced excessive volume is {a key} control on the flow dynamics including the flow velocity, depth, travel distance and impact area, {in turn affecting} the number of fatalities (Huggel et al., 2005; Evans et al., 2009; Le and Pitman, 2009; Dowling and Santi, 2014).
 The spatially varying erosion rates and entrainment processes are dependent on the geomorphological, lithological and mechanical conditions (Berger et al., 2011; Iverson et al., 2011; Reid et al., 2011; McCoy et al., 2012; Dietrich and Krautblatter, 2019).
A proper understanding of landslide erosion, entrainment and resulting increase in mass is a basic requirement for an appropriate modelling of landslide motion and its impact because the associated risk is directly related to the landslide
momentum. However, as mechanical controls of erosion and entrainment {are not} well understood yet, despite large efforts in the recent years, evolving volume, mobility and impact {forces} of landslides and debris flows are often quite improperly estimated (Dietrich and Krautblatter, 2019).
\\[3mm]
 Recently, there has been a rapid increase in the studies of erosion and
entrainment both in laboratory (Egashira et al., 2001; Fraccarollo and Capart, 2002; Iverson et al., 2011; de Haas et al., 2016),
and field scales (Cuomo et al., 2014, 2016; de Haas et al., 2020).
 Empirical (Takahashi and Kuang, 1986; Rickenmann et al., 2003; McDougall and Hungr, 2005; Chen et al., 2006; Le and Pitman, 2009)
and mechanical (Fraccarollo and Capart, 2002; Iverson, 2012)
erosion models have been developed. Although by nature mass transports are multi-phase phenomena (Pudasaini and Mergili, 2019), most erosion models only consider
effectively single-phase, 
 {or at most quasi two-phase} 
flows (Fraccarollo and Capart, 2002; McDougall and Hungr, 2005; Armanini et al., 2009; Le and Pitman, 2009; Iverson, 2012).
 {Different} numerical
models incorporating erosion
{have been proposed} (Le and Pitman, 2009; Iverson and Ouyang, 2015; McDougall and Hungr, 2005; Christen et al., 2010; Frank et al., 2015).
However, the erosion
rates presented and utilized therein are either not based on
physical principles, or {these} are physically incomplete, because they do not consider all the complex mechanical interactions between the materials in the flow and the erodible bed.
Moreover, these models are inconsistent, because they do not include the erosion-induced net momentum productions, and not all interactions across the erosion-interface are physically meaningfully considered (de Haas et al., 2020; Pudasaini and Fischer, 2020a).
{These physical shortcomings demand for a comprehensive and complete descriptions of the multi-phase erosion-entrainment process.} This will become clearer in the model development section.  
\\[3mm]
Erosion and deposition play an important role in mass transport and
evolution of the landscape (Huggel et al., 2005; Evans et al., 2009; Dietrich and Krautblatter, 2019; Mergili et al., 2020).
However, 
our understanding of these
processes is much below than needed to apply 
them to real events. 
Because of the complexity of the terrain and sporadic nature of the landslide events,
the time and cost demands for their field measurements are high. Yet, these measurements are only discrete. This limits the scope and utility (de Haas et al., 2020) of the available field data (Berger et al., 2011; Sch\"urch et al., 2011; McCoy et al., 2012; Theule et al., 2015;  Dietrich and Krautblatter, 2019).
Physics-based advanced and comprehensive models (Pudasaini and Fischer, 2020a), and sophisticated numerical simulations (Pudasaini and Mergili, 2019) can
overcome these limitations aiming to facilitate a more complete
understanding by investigating much wider aspects of the flow parameters,
erosion, mobility and deposition (Pudasaini and Krautblatter, 2021; Mergili et al., 2020).
\\[3mm]
 Pudasaini and Fischer (2020a) proposed a 
 process-based
erosion-deposition model for two-phase mass flows consisting of viscous fluid and solid particles, which, to a large extent, is capable of describing the complex erosive
phenomena commonly observed in landslides, avalanches and debris flows. Their mechanical erosion-rate models proved that the
effectively reduced friction (force) in erosion is equivalent to the momentum
production. This shows that erosion can enhance the mass flow mobility.
The importance of the Pudasaini and
Fischer (2020a) mechanical erosion model for two-phase mass flows 
{is}
widely realized in simulations of  
the real 
{catastrophic} multi-phase
events (Li et al., 2019; Qiao et al., 2019; Shen et al., 2019; Liu and He, 2020; Mergili et al., 2020; Liu et al, 2021).
These modelling approaches have clearly indicated the need of the mechanical erosion model 
in appropriately simulating the actual flow dynamics,
run-out, and deposition morphology based on the mechanical erosion rates
 and the erosion-induced momentum productions.
\\[3mm]
The Pudasaini and Fischer (2020a) two-phase model built a foundation for erosive mass flows by mechanically including the momentum production into the momentum balance equation. However, they did not include the inertia of the entrained mass and they could not present a clear mechanical condition for when and how the mobility of an erosive landslide will be enhanced or reduced and how to quantify it. So, their model appeared to be incomplete.
Extending the Pudasaini and Fischer (2020a) model, Pudasaini and Krautblatter (2021) addressed the important issue of erosion-induced landslide mobility by explicitly deriving mechanical conditions for the mobility of erosive landslides in terms of the erosion velocity of the mobilized bed material. 
They mechanically explained how and when erosive landslides enhance or reduce their mobility. 
This has been made possible by physically correctly considering the inertia and the momentum production of the erosive landslide. 
This model distinctly quantifies the mobility of an erosive landslide. 
Pudasaini and Krautblatter (2021) revealed that the erosion velocity determines the energy budget of an erosive landslide and provides an accurate description of mobility. They identified a novel mechanism of landslide-propulsion providing the erosion-thrust to the landslide. 
They constructed the mobility scaling
that precisely quantifies the contribution of erosion in landslide mobility.
They also derived a set of dynamical equations in which the momentum balance correctly includes the erosion-induced change in inertia and the momentum production, together called the net momentum production. Their model constitutes a foundation for physically meaningful simulation of landslide motion with erosion. However, their model is only for effectively-single phase solid-type materials in the flow and in the bed.
Also, the Pudasaini and Fischer (2020a) two-phase erosion model appeared to be incomplete, because it does not contain all natural interactions between phases across the erosion-interface, and requires fundamental enhancements to make it physically fully consistent and complete that follows the nature of erosive mass flows. This is what we achieve here with an innovative modelling approach. 
\\[3mm]
Based on the Pudasaini and Fischer (2020a) and Pudasaini and Krautblatter (2021) models, here, we develop a unified, consistent and complete mechanical erosion model for multi-phase mass flows.
This is the first-ever model to do so.
There are three major aspects of this contribution in relation to erosive multi-phase mass flows. ($i$) To physically correctly establish the jumps in shear stresses and momentum fluxes across the erosion-interface between the landslide and the bed substrate, and construct unified, comprehensive and consistent mechanical erosion rates for the solid and fluid phases. 
 We propose multi-phase mechanically-described interactive shear structures containing all the interactions between different phases on either side of the erosion-interface between the flowing landslide and the erodible bed substrate. 
 The sum of the solid and fluid erosion rates will be the total basal erosion rate, while the solid stresses include both the frictional and collisional stresses in an unified way. 
 The new erosion rate models are broad, well defined and well constrained. 
 ($ii$) To construct extensive and complete erosion-induced net momentum productions for both the solid and fluid phases. This is based on the physically correctly described novel complex erosion velocities of the mobilized particles and fluid from the basal substrate and the unified erosion rates. 
 ($iii$) To develop a realistic and comprehensive multi-phase mechanical erosion model that includes the novel, unified mechanical erosion rates, extended erosion velocities and the advanced net momentum productions into the mass and momentum balance equations. 
 As it includes different stress generating mechanisms; frictional, collisional and viscous; the new approach makes a complete description of the multi-phase erosive landslide by considering all the aspects associated with the erosion-induced momentum productions and the correct handling of the inertia while incorporating the net momentum productions. 
Moreover, the newly developed stress correction factor, erosive-shear-velocity, super-erosion-drift, and the erosion-matrix further highlight the importance of our modelling approach for the complex erosive multi-phase mass flows. 

\section{Model development for a unified multi-phase mechanical erosion rate}

The mechanical principle of erosion is based on the jump in shear stresses and the jump in momentum fluxes across the erosion interface between the landslide and the bed materials (Fraccarollo and Capart, 2002; Pudasaini and Fischer, 2020a). So, the major task here is to physically correctly construct these two jump quantities and relate them to a unified mechanical erosion rate for a multi-phase mass flow. However, first, we justify the real need of the new modelling philosophy. Then, we proceed with constructing the unified erosion rate model for multi-phase mass flows.

\subsection{Necessity of a unified and consistent multi-phase mechanical erosion rates,\\ and net momentum productions}

Pudasaini and Fischer (2020a) established a foundation of two-phase mechanical erosion model for mass flows. However, their model still requires a fundamental advancement in truly solving problems of erosive mass transports.
For several reasons, a novel unified multi-phase mechanical erosion model is required: 
The sum of solid and fluid erosion rates must be the total erosion rate of the erodible bed substrate. 
All the interactions between the solids and fluids in the landslide and the bed substrate must be considered.
All the shear resistances from the bed against all the applied shear stresses from the landslide must be mechanically explained, consistent and appropriate.
We must completely model the essentially complex erosion velocities (velocities of the eroded particles and fluid from the basal substrate) and produce the correct net momentum productions.
We must finally construct a comprehensive and unified mechanical model for multi-phase mass flows. 
These are explained below. To simplify the situation, consider the two-phase materials as an acceptable representation of multi-phase flows (Pudasaini, 2012) in both the landslide and the bed consisting of frictional as well as collisional solid particles, and viscous fluid.
 However, the derived model can be extended to multi-phase erosive flows.
\\[3mm]
 {\bf I. Consistency of the total basal erosion rate:} 
 Out of the total eroded material from the bed, the solid and fluid fractions must be consistently incorporated into the solid and fluid components in the moving material. However, this is a challenging problem that cannot be fixed with the existing two-phase mechanical erosion model (Pudasaini and Fischer, 2020a). As the solid and fluid erosion rates in Pudasaini and Fischer (2020a); the only existing two-phase erosion model; are developed independently, they do not include the basal solid and fluid volume fractions as multipliers. So, in technical applications, the values of solid and fluid erosion rates therein must be adjusted by choosing several parameters 
 such that the solid and fluid fractions of the eroded material is added in a desired way to the solid and fluid components in the moving mass. Due to this constraint, in real flow situations, the sum of these erosion rates do not realistically correspond to the eroded material from the bed. Moreover, as the fluid erosion rate in Pudasaini and Fischer (2020a) model is a function of the fluid velocity, for fast flows, the fluid erosion rate can be unrealistically high, and in turn, the solid erosion rate can be unrealistically low. However, often, the selected model parameters cannot adjust the erosion rates that correspond to the material composition in the erodible bed. So, natural erosion rates that are inline with the events cannot be obtained from the independent solid and fluid erosion rates developed by Pudasaini and Fischer (2020a). 
 This poses a great problem in consistent and appropriate selection of physical parameters appearing in their erosion rate models.
  This demands for a unified erosion rate model such that the sum of the derived solid and fluid erosion rates automatically satisfies the required natural criterion of the total erosion rate without any need of adjustment. We will achieve this here by first developing the total mechanical erosion rate for the mixture, that can be uniquely and legitimately split in to the solid and fluid erosion rates, which inherently contain the respective solid and fluid volume fractions of the bed material. Then, the total erosion rate is consistently and exactly obtained by summing up the solid and the fluid erosion rates, removing the great hurdle in the Pudasaini and Fischer (2020a) erosion rates. This is exactly what is needed in technical applications.
 \\[3mm]
 {\bf II. All interactions between solids and fluids in the landslide and bed substrate:} It is intuitively clear that there are four interactions between the solid and fluid phases in the landslide and the bed across the erosion-interface. These are: the solid-solid, solid-fluid, fluid-solid and fluid-fluid interactions. However, the Pudasaini and Fischer (2020a) two-phase erosion model only considers the direct phase-phase interactions, i.e., solid-solid and fluid-fluid interactions between the landslide and the bed, but ignores the cross-phase interactions, i.e., the solid-fluid, fluid-solid interactions. These interactions, though quite natural, are not yet recognized. This severely limits the applicability of existing models in simulating real multi-phase erosive events. We overcome this problem here by considering all phase-phase and cross-phase interactions. 
 \\[3mm]
 {\bf III. Mechanically explained bed shear resistances against applied shear stresses from landslide:} 
 As there are four interactions between the solid and fluid phases, there must be four basal shear resistances against the four applied shear stresses from the landslide. As only solid-solid and fluid-fluid interactions are considered by Pudasaini and Fischer (2020a), there are two main shortcomings in their erosion rate modelling. First, the solid-solid shear stresses are modelled by applying the often used Coulomb-type frictional law. However, for the fluid-fluid interactions, classical Chezy-type friction law was applied, which as we will see later, is not appropriate. We eliminate this problem here by developing novel, mechanically appropriate fluid-fluid shear stress models. Second, the true solid-fluid and fluid-solid interactions were ignored by Pudasaini and Fischer (2020a). Nevertheless, these interactions are substantial and appear to be quite complex than our present state of knowledge. So, based on the Chezy-type friction law, here, we present largely novel, mechanical shear stress structure for the solid-fluid interaction.
 \\[3mm]
 {\bf IV. Complementing the frictional stress with the collisional stress for solid particles:} 
 For solid particles, we combine the Coulomb-type frictional law with the Bagnold-type collisional law, whereas the collisional stress includes some novel considerations. We also suggest some criteria to dynamically characterize the frictional and collisional flow regimes.
 \\[3mm]
 {\bf V. Complex erosion velocities and correct multi-phase net momentum productions:} One of the most astonishing facts emerges here while dealing with the erosion velocities of the mobilized bed materials by the flow. As there are four phase-phase and cross-phase interactions between the solid and fluid materials in the landslide and the bed, erosion velocities for both the solid and fluid take extensive and complex forms. Consider the erosion velocity of a solid particle at the bed. This solid particle is pushed (sheared) and then mobilized by both the solid and  fluid in the flow. However, Pudasaini and Fischer (2020a) considered the mobilization of the basal solid particle only by the solid from the flow, but ignored the other component of mobilization of this solid particle by the fluid in the landslide (mixture). The same is true for the erosion velocity of the fluid molecule at the bed. With these novel realizations, 
 we constructed mechanically complex, comprehensive solid and fluid erosion velocities by considering all the mobilization components as applied by both the solid and fluid from the flow to the solid and fluid in the erodible bed (mixture). These will have huge implications in correctly describing net momentum productions, because the net momentum productions are functions of erosion velocities. This is very important, because, as we will see later, the net momentum production entirely controls the dynamics, mobility, impact energy and the deposition morphology of the mass transport. Moreover, the Pudasaini and Krautblatter (2021) erosion-induced landslide mobility model is only for effectively single-phase flows and only considers the single-phase net momentum production. However, as the real events are of multiphase nature, we must consistently extend the single-phase net momentum production to multi-phase net momentum productions, which however turn out to be quite elaborate.  
 \\[3mm]
 {\bf VI. Comprehensive multi-phase erosion model:} Limitations of the existing two-phase erosion model as exposed above in {\bf I} - {\bf V} demand for a consistent and complete multi-phase mechanical erosion model. This is achieved here by constructing a novel, unified mechanical erosion rates and momentum production models and embedding them into the dynamical model equations (mass and momentum balances). This paves the way for the legitimate use of the developed erosion models in applications, a monumental step forward in complex multi-phase mass flow simulations. 

\subsection{Definition of variables and parameters}

For simplicity, first we consider a two-phase landslide (flow material) and a bed morphology
consisting of viscous fluids and solid particles of different physical,
mechanical and geometrical properties. Later we discuss
how the derived models can be extended to multi-phase
debris mixture and the erodible bed substrate. 
The two-phase mixtures with solid and fluid phases are indicated by the subscripts $_s$ and $_f$. Let the physical and
mechanical parameters and flow dynamical variables in the landslide mixture be
denoted by the superscript $^m$ and the erodible material in the bed be denoted by the superscript $^b$,
respectively. 
Let ${\bf u} = (u,v)$ be the flow velocity, where $u$ and $v$ are the downslope ($x$) and cross slope ($y$) components.
Furthermore, let $\alpha, \rho, u, \mu$ and $\tau$ denote the
volume fraction, density, velocity, friction coefficient and the shear stress. Then, the mixture
densities, velocities and the shear stresses on either sides of the erosion
interface are given by (Pudasaini and Fischer, 2020b):
%\begin{align}
\begin{equation}
\rho^m = \alpha_s^m \rho_s^m + \alpha_f^m \rho_f^m,\,\,\,\,\,
\rho^m u^m = \alpha_s^m \rho_s^m u_s^m + \alpha_f^m \rho_f^m u_f^m,\,\,\,\,\, 
\tau^m = \tau_s^m + \tau_f^m,
\label{Eqn_1}
\end{equation}
\begin{equation}
\hspace{-1.8cm}
\rho^b = \alpha_s^b \rho_s^b + \alpha_f^b \rho_f^b,\,\,\,\,\,
\rho^b u^b = \alpha_s^b \rho_s^b u_s^b + \alpha_f^b \rho_f^b u_f^b,\,\,\,\,\,
\tau^b = \tau_s^b + \tau_f^b,
\label{Eqn_2}
\end{equation}
%\end{align}
where the hold-ups $\alpha_s^m  + \alpha_f^m  =1$ and $\alpha_s^b  +
\alpha_f^b  =1$ are satisfied.
The quantities without subscripts are the total quantities associated with the mixtures, either in the flow $\left ( \rho^m, u^m, \tau^m\right )$ or in the bed $\left ( \rho^b, u^b, \tau^b\right )$. 
\\[3mm]
{To connect the velocities at the lower level (denoted by $_l$) of the flow (landslide) and the erodible substrate to the mean flow velocities, following Pudasaini and Fischer (2020a), we define the following extended relations,
\begin{eqnarray}
\begin{array}{lll}
&&u_l^m = \lambda^m_l u^m,\,  u^b = \lambda^b u^m; \,\, 
u_{s_l}^m = \lambda_{s_l}^m u_s^m,\, u_{f_l}^m = \lambda_{f_l}^m u_f^m; \,\,
u_{{sf}_l}^m = \lambda_{{sf}_l}^m u_s^m,\, u_{{fs}_l}^m = \lambda_{{fs}_l}^m u_f^m;\\[3mm]
&&u_{ss}^b = \lambda_{ss}^b u_s^m,\, u_{ff}^b = \lambda_{ff}^b u_f^m;\,\,
u_{fs}^b = \lambda_{fs}^b u_f^m,\, u_{sf}^b = \lambda_{sf}^b u_s^m,
\label{Eqn_5}
\end{array}
\end{eqnarray}
where,  $\lambda_l^m , \lambda^b $; $\lambda^m_{s_l}, \lambda^m_{f_l}; 
\lambda_{{sf}_l}^m, \lambda_{{fs}_l}^m;
\lambda_{ss}^b, \lambda_{ff}^b;
\lambda_{fs}^b, \lambda_{sf}^b$
are called the erosion drifts,
and their mechanical closures are determined at Section 3.1.3 by the erosion-drift equations. 
These erosion drifts are positive numbers, and are bounded from above by unity.
The relations in (\ref{Eqn_5}) are needed while deriving erosion (basal) velocities (see below), shear stresses (Section 2.4), erosion rates or mass productions (Section 3) and momentum productions (Section 4).
The origin and use of the double subscripts $\left ( ss, sf, fs, ff\right )$ for the phase interactions has been explained below and at Section 2.3 and Section 2.4. 
In (\ref{Eqn_5}) $u_{{sf}_l}^m$ is the velocity of the solid at the base of the landslide as it is in contact with the fluid at the bed. Similar definition applies for $u_{{fs}_l}^m$.
Although there are ten erosion drifts in (\ref{Eqn_5}), which we had to define for formal reasons, only few of them are independent (and needed) which can be modelled or described relatively easily as functions of known relevant bed inertial numbers (associated effective density ratios, Pudasaini and Krautblatter, 2021), as we will see at Section 3.1.3.
Because, only very few of the erosion drifts will be needed in the final model equations, and simple mechanical closures are developed for all of them, there is absolutely no agnst.
  Note that $\lambda_l^m , \lambda^b$ are associated with the total mixtures, and $\lambda^m_{s_l}, \lambda^m_{f_l}$, 
 $\lambda_{{sf}_l}^m, \lambda_{{fs}_l}^m$,
  $\lambda_{ss}^b, \lambda_{ff}^b$ and $\lambda_{fs}^b, \lambda_{sf}^b$ are associated with the solid and fluid phases (separately, or in combination) in the landslide and bed. These will be clearer in due courses.
  In Pudasaini and Fischer (2020a) only $\lambda^m_{s_l}, \lambda^m_{f_l}; \lambda_{ss}^b, \lambda_{ff}^b$ appear, but in reduced form, without any cross-phase interactions across the erosion-interface, limiting the applicability of their model.

\subsection{Solid and fluid erosion velocities}

In (\ref{Eqn_1}), (\ref{Eqn_2}) and (\ref{Eqn_5}) all the velocities $u$ with the superscript $^b$ are the erosion velocities, i.e., they are the velocities of the eroded materials, or are associated with the erosion velocities of the mobilized materials from the bed. 
For mixture materials, however, the erosion velocities of the solid particle and the fluid molecule at the bed are complex. 
There are four different interactions between the particles and fluid in the flow and the bed. For this reason, we introduced four double subscripts for the erosion velocities, similarly for the shear stresses and shear resistances (Section 2.4), namely the solid-solid $(ss)$, solid-fluid $(sf)$, fluid-solid $(fs)$, and fluid-fluid $(ff)$ interactions.
This is so, because the solid particle at the bed is sheared (pushed) by both the solid and fluid at the base of the flow (landslide). 
It means, the solid erosion velocity $u_s^b$ has two components. It is jointly induced by the pushing of the solid from the landslide, which we denote by $u_{ss}^b$, and by the pushing of the fluid from the landslide, which we denote by $u_{fs}^b$, respectively.
Similarly, the fluid molecule at the bed is also sheared~(pushed) by both the solid and fluid at the base of the flow which are denoted by the components $u^b_{sf}$ and $u^b_{ff}$. 
However, the basal velocity components $u^b_{ss}$ and $u^b_{fs}$ for solid are induced by the solid and fluid fractions $\alpha_s^m$ and $\alpha_f^m$ in the flowing mixture, because the solid particles at the bed are pushed by both the solid and fluid at the base of the flow. 
This is an important realization.
The same is true for the fluid erosion velocity components $u^b_{sf}$ and $u^b_{ff}$ 
as the fluid molecules at the bed are pushed by both the solid and fluid at the base of the flow. 
So, as invented here, the solid and fluid erosion velocities take their complex forms:
\begin{eqnarray}
\begin{array}{lll}
\displaystyle{u_s^b = \alpha_s^m u_{ss}^b + \alpha_f^m u_{fs}^b = \alpha_s^m \lambda_{ss}^b u_s^m + \alpha_f^m \lambda_{fs}^b u_f^m},\\[3mm]  
\displaystyle{u_f^b = \alpha_s^m u_{sf}^b + \alpha_f^m u_{ff}^b = \alpha_s^m \lambda_{sf}^b u_s^m + \alpha_f^m \lambda_{ff}^b u_f^m}, 
\label{Eqn_1NNN}
\end{array}
\end{eqnarray}
where the erosion drifts are employed from (\ref{Eqn_5}). 
New drift parameters (or functions) $\lambda_{fs}^b$ and $\lambda_{sf}^b$ appear in (\ref{Eqn_5}) and (\ref{Eqn_1NNN}). We call them the cross-phase erosion drifts. 
Similarly, $\lambda_{ss}^b$ and $\lambda_{ff}^b$ are called the direct phase-erosion drifts.
Once the erosion drifts $\left ( \lambda_{ss}^b, \lambda_{fs}^b\right )$ and $\left ( \lambda_{sf}^b, \lambda_{ff}^b\right )$ are determined at Section 3.1.3, the basal erosion velocities in (\ref{Eqn_1NNN}) are closed, because the phase volume fractions $\left (\alpha_s^m, \alpha_f^m \right)$ and velocities $\left (u_s^m, u_f^m \right)$ are the state variables in the mass and momentum balance equations, and are presented at Section 7.

\subsection{Shear stresses}

As the erosion rate depends on the shear stress jump $\tau^m - \tau^b$ across the erosion interface, first, we need to describe all the shear stresses and their jumps across the erosion-interface.  
In the Pudasaini and Fischer (2020a) two-phase erosion model, only the direct solid-solid and fluid-fluid interactions between the landslide and bed materials were considered. This is incomplete and partly inconsistent. Incomplete, because, in general, there are four phase-interactions across the erosion-interface. Inconsistent, because the fluid-fluid interactions $(ff)$ must follow some other mechanical principles, but not the Chezy-type frictions. Moreover, there are solid-fluid $(sf)$ and fluid-solid $(fs)$ interactions which were not considered in the Pudasaini and Fischer (2020a) model. This will be clearer below. 
\\[3mm]
As mentioned above, consider the two-phase mixture landslide $\left( ^m\right )$ and bed material $\left ( ^b \right)$, both consisting of solid ($s$) and fluid ($f$) phases.
When both the landslide (flow) and the erodible bed are composed of two-phase (or multi-phase) materials of different physical properties, the shear stresses are non-trivial and can become very complicated. 
There are two types of interactions. Solid from the flow interacting with the solid and fluid in the bed, and fluid from the flow interacting with the solid and fluid in the bed. These interactions are discussed below in detail.
\\[3mm]
{\bf I. Solid in the landslide interacting with solid and fluid in the bed:} The solid particles in the landslide (flow) are in contact either with the solid or with the fluid in the bed. This introduces two interactions: solid from the flow interacting with the solid in the bed ($ss$: given by $ \alpha_f^b\left [ \tau_{ss}^m - \tau_{ss}^b \right ] $ on the left hand side in (\ref{Eqn_1N}) below), and solid from the flow interacting with the fluid in the bed ($sf$: given by $ \alpha_s^b\left [ \tau_{sf}^m - \tau_{sf}^b \right ]$ on the left hand side in (\ref{Eqn_1N}) below). However, as the solid phase in the mixture talks with the solid in the bed, it tells that its solid stress has to be uniquely distributed between the fluid $\left (\alpha_f^b\right )$ and the solid $\left (\alpha_s^b\right )$ phases in the bed, and mechanically they must agree. 
This is the reason why these factors appear in (\ref{Eqn_1N}) below.
This results in two different (solid-solid, solid-fluid) interactions 
\begin{equation}
\alpha_f^b\left [ \tau_{ss}^m - \tau_{ss}^b \right ] + 
\alpha_s^b\left [ \tau_{sf}^m - \tau_{sf}^b \right ]
\!=\!\alpha_f^b\left [ \tau_{ss}^m - \tau_{ss}^b \right ] + 
\alpha_s^b\left [ \tau_{ss}^m - \tau_{sf}^b \right ]
\!=\! \left[\alpha_f^b + \alpha_{s}^b \right] \tau_{ss}^m - \alpha_f^b \tau_{ss}^b - \alpha_s^b \tau_{sf}^b
\!=\! \tau_{ss}^m - \alpha_f^b \tau_{ss}^b - \alpha_s^b \tau_{sf}^b, 
\label{Eqn_1N}
\end{equation}
where the double suffix $ss$ in $\tau_{ss}^m$ means the shear stress is applied by the solid in the landslide mixture (first $s$ in $ss$) to the solid in the bed (second $s$ in $ss$), and $ss$ in $\tau_{ss}^b$ means the shear resistance by the solid from the bed (second $s$ in $ss$) against the applied shear stress by the solid from the landslide (first $s$ in $ss$). 
Similarly, the suffix $sf$ in $\tau_{sf}^m$ means the shear stress is applied by the solid in the landslide mixture (first $s$ in $sf$) to the fluid in the bed (second $f$ in $sf$), 
and the suffix $sf$ in $\tau_{sf}^b$ means the shear resistance by the fluid from the bed (second $f$ in $sf$) against the applied shear stress by the solid from the landslide (first $s$ in $sf$). 
Since the solid material in the mixture always behaves as the solid whether it interacts with the solid or fluid in the bed, $\tau_{sf}^m = \tau_{ss}^m$. This has been employed in (\ref{Eqn_1N}). 
In the sequel, other similar double suffix notations will have analogous meanings. So, the double suffix represents the type of interactions and motions between the flow and the bed.  
\\[3mm]
{\bf II. Fluid in the landslide interacting with solid and fluid in the bed:} Fluid molecules in the flow interact either with the solid or with the fluid in the bed. This introduces two further interactions: fluid from the flow interacting with the solid in the bed ($fs$), and fluid from the flow interacting with the fluid in the bed ($ff$). Again, as the fluid phase in the mixture communicates with the solid in the bed, it tells that this fluid stress has to be distributed between the solid $\left (\alpha_s^b\right )$ and the fluid $\left (\alpha_f^b\right )$ phases in the bed. This results in two different additional (fluid-solid, fluid-fluid) interactions 
\begin{equation}
\alpha_s^b\left [ \tau_{fs}^m - \tau_{fs}^b \right ] + 
\alpha_f^b\left [ {\tau}_{ff}^m - \tau_{ff}^b \right ]
=\alpha_s^b\left [ \tau_{fs}^m - \tau_{ss}^b \right ] + 
\alpha_f^b\left [ {\tau}_{ff}^m - \tau_{ff}^b \right ].
\label{Eqn_2N}
\end{equation}
In (\ref{Eqn_2N}) too, on the left hand side, $\tau_{fs}^b$ indicates the shear resistance of the basal solid material against the applied fluid shear stress from the flow, $\tau_{fs}^m$. Similarly, $\tau_{ff}^b$ indicates the shear resistance of the basal fluid material against the applied fluid shear stress from the flow, $\tau_{ff}^m$. 
Since the basal resistance from the solid in the bed is the same whether it is against solid or fluid from the mixture, $\tau_{fs}^b = \tau_{ss}^b$, which is applied on the right hand side of (\ref{Eqn_2N}). 
However, note that due to the responses from the essentially distinct solid and fluid materials, there are fundamental differences between (\ref{Eqn_1N}) and (\ref{Eqn_2N}). The way the fluid in the mixture applies the shear stress to the solid in the bed $\left ( \tau_{fs}^m\right)$ and the fluid in the bed $\left ( \tau_{ff}^m\right)$ are different. 
Due to this, unlike (\ref{Eqn_1N}), (\ref{Eqn_2N}) could not be reduced further.
These need to be modelled separately and carefully. This will be clearer at Section 2.4.2 and Section 2.4.3.
\\[3mm]
In (\ref{Eqn_1N}) and (\ref{Eqn_2N}), the $+$ and $-$ signs, respectively, are associated with the applied shear stresses from the flowing mixture and the shear resistances from the erodible bed material. By adding (\ref{Eqn_1N}) and (\ref{Eqn_2N}) we obtain the net shear stress of the system (applied shear stresses from the mixture flow minus shear resistances from the bed):
\begin{equation}
\tau_{ss}^m - \alpha_f^b \tau_{ss}^b - \alpha_s^b \tau_{sf}^b
+\alpha_s^b\left [ \tau_{fs}^m - \tau_{ss}^b \right ] + 
\alpha_f^b\left [ {\tau}_{ff}^m - \tau_{ff}^b \right ]
= \left[\tau_{ss}^m + \alpha_s^b \tau_{fs}^m + \alpha_f^b {\tau}_{ff}^m\right ]
- \left[\tau_{ss}^b + \alpha_s^b \tau_{sf}^b + \alpha_f^b \tau_{ff}^b \right ].
\label{Eqn_3N0}
\end{equation}
So, the jump in the shear stress $\tau^m - \tau^b$ (or the net shear stress) across the erosion-interface can be written as:
\begin{equation}
\tau^m - \tau^b 
= \left[\tau_{ss}^m + \alpha_s^b \tau_{fs}^m + \alpha_f^b {\tau}_{ff}^m\right ]
- \left[\tau_{ss}^b + \alpha_s^b \tau_{sf}^b + \alpha_f^b \tau_{ff}^b \right ],
\label{Eqn_3N}
\end{equation}
where $\tau^m$ and $\tau^b$ are the total shear stress applied by the flow and the shear resistance by the bed, respectively.
The shear stress jump in (\ref{Eqn_3N}) can also be written as:
\begin{equation}
\tau^m - \tau^b 
= \left[\tau_{ss}^m - \tau_{ss}^b\right ]
+ \alpha_s^b\left[  \tau_{fs}^m - \tau_{sf}^b\right ]
+ \alpha_f^b\left[  \tau_{ff}^m - \tau_{ff}^b\right ].
\label{Eqn_3al}
\end{equation}
{\bf Importance of the shear stress jump in (\ref{Eqn_3N}):}
There are several important aspects associated with the net shear stress jump of the system as seen in (\ref{Eqn_3N}). The uniqueness and legitimacy of the elegant cross-couplings in (\ref{Eqn_1N}) and (\ref{Eqn_2N}) have been discussed at Section 2.5 and Section 6.1.
\\[3mm]
{\bf A.} The first terms on the right hand side in the square brackets in (\ref{Eqn_3N}), i.e., $\tau_{ss}^m$ and $\tau_{ss}^b$ are due to the shear stress applied by the solid in the mixture and the resistance from the solid in the bed, respectively. As discussed at Section 2.4.1, these terms satisfy the Coulomb-type frictional rheologies and Bagnold-type collisional rheologies because they originate from solid-solid interactions. 
\\[3mm]
{\bf B.} The second terms on the right hand side in the square brackets in (\ref{Eqn_3N}), i.e., $\alpha_s^b \tau_{fs}^m$ and $\alpha_s^b \tau_{sf}^b$ are the shear stress applied by the fluid in the landslide against the solid in the bed, and the shear resistance by the fluid in the bed against the applied shear stress by the solid from the landslide. As these terms emerge from the fluid-solid and solid-fluid interactions, they satisfy the Chezy-type frictional rheologies, but with some crucial amendments revealed here. Models for these shear stresses are presented at Section 2.4.2.
\\[3mm]
{\bf C.} However, the last terms on the right hand side in the square brackets in (\ref{Eqn_3N}), i.e., $\alpha_f^b {\tau}_{ff}^m$ and $\alpha_f^b \tau_{ff}^b$ are the fluid-fluid interactions between the fluid in the landslide (applied shear stress) and the fluid in the bed (shear resistance). These need to be described in a new way as there exist no model for erosive mass flows for such interactions. Models for these shear stresses are presented at Section 2.4.3.
\\[3mm]
The major task lies in modelling each of the shear stresses in (\ref{Eqn_3N}), which we deal with below for all six components (three shear stresses applied by the landslide material, plus terms in the first square bracket, and three shear resistances from the erodible bed, minus terms in the second square bracket).

\subsubsection{Coulomb-type and Bagnold-type shear stresses for solid-solid interactions}

{\bf A. Coulomb-type frictional stresses:}
 Following the classical approaches (Pudasaini and Fischer, 2020a), the solid phase
shear stresses in (\ref{Eqn_3N}) induced by the frictional solid-solid interactions between 
the landslide and the erodible basal surface are modelled by the Coulomb-type shear stresses, and are given by
\begin{eqnarray}
\begin{array}{lll}
\tau_{C, ss}^m = \left ( 1 - \gamma^m\right ) \rho_s^m g^z h\mu_s^m\alpha_s^m,\\[3mm]
\tau_{C, ss}^b = \left ( 1 - \gamma^b\right ) \rho_s^b g^z h \mu_s^b\alpha_s^b,
\label{Eqn_1NN}
\end{array}
\end{eqnarray}
where $\gamma^m = \rho_f^m/\rho_s^m$ and $\gamma^b = \rho_f^b/\rho_s^b$ are the density ratios, and the terms  $\left ( 1 - \gamma^m\right )$ and $\left ( 1 - \gamma^b\right )$ emerge due to the buoyancy reduced normal loads of the respective solid particles in the landslide and the bed mixtures. Moreover, $h$ is the total flow depth (solid plus fluid); $\mu_s^m = \tan\delta_s^m$ and $\mu_s^b = \tan\delta_s^b$ are the friction coefficients corresponding to the friction angles $\delta_s^m$ and $\delta_s^b$, and $g^z$ is the component of gravitational acceleration in the direction normal to the slope.
Out of the six shear stresses in (\ref{Eqn_3N}), these Coulomb-type shear stresses for solid-solid interactions appeared to be the most simple to model. 
\\[3mm]
{\bf B. Bagnold-type collisional stresses:}
When the granular flows also develop significant to dominant collisional stresses, we should consider these interactions. As these complex processes involve quite a bit of careful additional work, the models are constructed separately later in Section 5 for both the flow and the bed. In what follows, the resulting comprehensive total shear stresses $\tau_{ss}^m = \tau_{C, ss}^m + \tau_{B, ss}^m$ and $\tau_{ss}^b = \tau_{C, ss}^b + \tau_{B, ss}^b$, where $B$ in $\tau_{B}$ indicates the Bagnold-type collisional stress, in (\ref{Eqn_Col6}) or (\ref{Eqn_Col6_a}), and (\ref{Eqn_Col11}) or (\ref{Eqn_Col11_a}) are implemented while developing the unified erosion model.

\subsubsection{Novel Chezy-type shear stresses for fluid-solid and solid-fluid interactions}

Following the classical approaches (Fraccarollo and Capart, 2002), the fluid-solid and solid-fluid
shear stresses in (\ref{Eqn_3N})
between the landslide and the erodible basal surface  are described by the Chezy-type shear stresses (Pudasaini and Fischer, 2020a): 
\begin{eqnarray}
\begin{array}{lll}
\tau_{fs}^m =   C_f^m \rho_f^m \left ( u_{f_l}^m\right )^2\alpha_f^m h/H
=   C_f^m \rho_f^m \left ( \lambda_{f_l}^m u_f^m\right )^2\alpha_f^m h/H,
\\[3mm]
\tau_{sf}^b =   C_f^b \rho_f^b \left ( u_{sf}^b\right )^2\alpha_f^b h/H
= C_f^b \rho_f^b \left ( \lambda_{sf}^b u_s^m\right )^2\alpha_f^b h/H,
\label{Eqn_2NN}
\end{array}
\end{eqnarray}
where $C_f^m$ and $C_f^b$ are the Chezy-friction coefficients (Chow, 1959), and the erosion drifts $\lambda_{f_l}^m$ and $\lambda_{sf}^b$ are defined at Section 2.2 and their mechanical closures are presented at Section 3.1.3.
It is very important to observe that, the fluid-type resistance from the bottom material against the shear stress from the solid in the landslide, i.e., $\tau_{sf}^b$, contains a cross-velocity expression $u_{sf}^b = \lambda_{sf}^b u_s^m$. As mentioned at Section 2.3, the fluid motion in the bed for this situation is partly determined by the solid motion at the base, but not by the fluid motion at the base of the landslide. This is a special circumstance encountered and invented here due to the solid-fluid interactions, but not realized previously in erosive mass transports. It poses a challenge, yet, equally unravels the physical situation associated with such an interaction.
This is in contrast to the usual mechanism in $\tau_{fs}^m $ which includes the fluid motion at the base of the landslide material. This is natural.
In this respect, the structure of $\tau_{sf}^b$ is a fundamentally new understanding over the previous fluid-type bed shear resistance against the solid-type shear stress from the landslide in Pudasaini and Fischer (2020a), in which, simply $u_{ff}^b = \lambda_{ff}^b u_f^m$ was used, which for the mixture interactions, appeared to be physically inappropriate. This is a notable aspect.
Also, contrary to the classical Chezy-type law, because of the mixtures in our consideration, as seen in (\ref{Eqn_2N}) and (\ref{Eqn_2NN}), the bed solid and landslide fluid volume fractions appear both along with the shear stresses $\tau^m_{fs}$ and $\tau^b_{sf}$.

\subsubsection{Novel shear stress descriptions for fluid-fluid interactions}

Although there are some crucial innovative aspects in the Chezy-type shear stresses for fluid-solid and solid-fluid interactions in (\ref{Eqn_2NN}), structurally the Coulomb-type and Chezy-type shear stresses are known and have been applied previously for the erosive mass flows (Fraccarollo and Capart, 2002; Pudasaini and Fischer, 2020a). However, for the erosive landslide, the fluid-fluid interactions are not yet known, and we need to consider (construct) some new physically-based models. 
Assuming that, momentarily before the erosion takes place, the fluid in the bed material is stationary, or its motion is negligible (this restriction can be lifted) as compared to the motion of the fluid in the landslide. 
For shear stresses that act at the interface between the viscous fluid in the landslide and the viscous fluid in the bed, pioneering work by Beavers and Joseph (1967) provides a basis for modelling such processes (Jones, 1973). We utilize this idea to model shear stresses between the fluid in the landslide mixture (as in a porous media) and the fluid in the mixture bed (also behaving as a porous media) and vice versa:
\begin{eqnarray}
\begin{array}{lll}
\displaystyle{{\tau}_{ff}^m  
                  = \eta_f^m \frac{\alpha_{_{BJ}}^b}{\sqrt{\mathcal K^b}} u_{f_l}^m
                  = \eta_f^m \frac{\alpha_{_{BJ}}^b}{\sqrt{\mathcal K^b}} \lambda_{f_l}^m u_{f}^m
                  = \eta_f^m \lambda_{f_l}^m\frac{\alpha_{_{BJ}}^b}{\sqrt{\mathcal K^b}} u_{f}^m,}\\[5mm]
\displaystyle{\tau_{ff}^b       
                  = \eta_f^b \frac{\alpha_{_{BJ}}^m}{\sqrt{\mathcal K^m}}u_{ff}^b
                  = \eta_f^b \frac{\alpha_{_{BJ}}^m}{\sqrt{\mathcal K^m}} \lambda_{ff}^b u_{f}^m
                  = \eta_f^b \lambda_{ff}^b\frac{\alpha_{_{BJ}}^m}{\sqrt{\mathcal K^m}} u_{f}^m,} 
\label{Eqn_3NN}
\end{array}
\end{eqnarray}
where, ${\alpha_{_{BJ}}}$ are dimensionless numbers, with typical values in $[0.7, 4.0]$ that depend on the physical parameters characterizing the structure of the landslide and bed materials (Beavres and Joseph, 1967), but independent of fluid viscosities on either side of the interface, and $\left ({\mathcal K^m}, {\mathcal K^b}\right)$ are the permeabilities of the landslide and the bed materials, respectively, which are small values, but depending on the material type, can vary several orders of magnitudes. 
Permeabilities are assumed to be given quantities rather than unknown parameters.
Furthermore, the erosion drifts $\lambda_{f_l}^m$ and $\lambda_{ff}^b$ are defined at Section 2.2 and are given at Section 3.1.3.
Shear stresses in (\ref{Eqn_3NN}) are distinguished by the fluid viscosities and velocities on their respective sides, however, with the permeability of the bed and the permeability of the landslide materials on opposite sides of the erosion-interface. 
Moreover, the shear stress $\tau_{ff}^b$ contains a component $u_{ff}^b$ of the bed fluid (erosion) velocity $u_f^b$ induced by the fluid velocity in the landslide $\left (u_{f}^m\right)$ which complicates the situation, but the model for this has been developed in (\ref{Eqn_1NNN}). This has been applied in (\ref{Eqn_3NN}). 
\\[3mm]
The relations in (\ref{Eqn_3NN}) are based on the hypothesis of slip boundary condition and is developed with a boundary layer approach (Nield, 2009). 
Also, note that in contrast to the Beavers-Joseph-type law, because of the mixtures in our consideration, as seen in (\ref{Eqn_3N}), the bed fluid volume fraction appears along with these shear stresses, and both sides of the interface contain solid particles, so forming porous material of different physical properties. 
As $\sqrt{\mathcal K^b}$ has the dimension of length, $\lambda_{f_l}^m \alpha_{BJ}^b u_{f}^m/\sqrt{\mathcal K^b}$ can be perceived as the rate of strain, which when multiplied with the respective viscosity $\eta_f^m$ results in the usual structure of the viscous shear stress at the landslide base: ${\tau}_{ff}^m \approx \eta_f^m \left [\partial u_{f_l}^m/\partial z\right ]_{z = b}$, where $z$ is the flow depth. The same applies to $\tau_{ff}^b$. This reveals the origin, and justifies the physical mechanisms, of the interfacial shear stresses in (\ref{Eqn_3NN}).
\\[3mm]
{\bf Simplified fluid-fluid interactions:} The expressions in (\ref{Eqn_3NN}) can further be simplified. Following Chandesris and Jamet (2006), if $\alpha_{_{BJ}}^b$ may be written as the ratio of the fluid viscosities in the bed and the landslide $\alpha_{_{BJ}}^b = \sqrt{\eta_f^b/\eta_f^m}$, similarly $\alpha_{_{BJ}}^m = \sqrt{\eta_f^m/\eta_f^b}$, then (\ref{Eqn_3NN}) can be reduced to yield:
\begin{equation}
{\tau}_{ff}^m  = \eta_e^{P} \frac{\lambda_{f_l}^m u_{f}^m}{\sqrt{\mathcal K^b}}, \,\,\,\,\,
\tau_{ff}^b       = \eta_e^{P} \frac{\lambda_{ff}^b u_{f}^m}{\sqrt{\mathcal K^m}}, 
\label{Eqn_4NN}
\end{equation}
where the geometric mean $\eta_e^{P} = \sqrt{\eta_f^m \eta_f^b}$ is the (net) effective system viscosity at the erosion-interface, which we call the erosive product viscosity, or simply the P-viscosity. 
So, the fluid shear stresses at the erosion-interface are dependent on the common P-viscosity, and the respective fluid velocities together with the drifts on their own sides, but inversely associated with permeabilities on opposite (resistive) sides of the landslide-bed interface.
\\[3mm]
To sum up, the jump in the shear stresses across the erosion-interface is given by (\ref{Eqn_3N}) together with the shear stress models developed in (\ref{Eqn_1NN})-(\ref{Eqn_3NN}).

\subsection{Reduced net shear stresses}

Depending on the composition of the flowing landslide and the bed mixture and the significance of the relevant shear stresses, amazingly as we figure out here, we will have eight different reduced (indicated by $_r$) net shear stresses $\left (\tau^{net}_r = \tau^m_r - \tau^b_r\right)$ of the system (shear stress jump across the erosion-interface). This also indicates the complexity associated with the two-phase erosive landslide, which have not been recognized in the landslide research as such large number of interactions were not identified previously. However, these need to be realized correctly while distributing the shear stresses among the respective components, in both the flow and bed materials as the landslide interacts with the bed. This is important. These are analyzed below.
\\[3mm]
{\bf I. Solid-type materials:} If both the landslide and bed materials are only of solid-type (dry landslide entraining dry bed), then (\ref{Eqn_3N}) directly reduces to the simple expression: 
\begin{equation}
\tau^m_r - \tau^b_r = \tau_{ss}^m - \tau_{ss}^b = \tau_{s}^m - \tau_{s}^b. 
\label{Eqn_4N}
\end{equation}
Since there are only solid materials, there are no cross-interactions, so, we used $s$ for $ss$. This recognition will also be applied below for those shear stresses when some or all cross-interactions can be ignored.  
\\[3mm]
{\bf II. Fluid-type materials:} If both the flow and bed materials are only of fluid-type (fluid flow entraining fluid material, e.g., flood entraining river, lake or reservoir fluid), then (\ref{Eqn_3N}) reduces to the simple expression: 
\begin{equation}
\tau^m_r - \tau^b_r 
= \alpha_f^b \left [ {\tau}_{ff}^m- \tau_{ff}^b \right ] = {\tau}_{ff}^m- \tau_{ff}^b = {\tau}_{f}^m- \tau_{f}^b,
\label{Eqn_5N}
\end{equation}
because, there is no solid means all other solid and cross-phase contributions do not exist and $\alpha_f^b = 1$, and $f$ is used instead of $ff$ as there are no cross-interactions. However, note that, in this situation, we should use simple fluid shear stress because there are no solid particles to form matrices constituting porous medium. Yet, in general, only fluid-fluid interaction is less likely to take place in erosive mass flows.
\\[3mm]
{\bf III. Solid-type landslide and fluid-type bed:} If the landslide is a solid-type material and the bed is a fluid-type material (dry landslide entraining river, lake or reservoir fluid), then the landslide solid stress does not need to be distributed but only applied to the fluid in the bed.
With these realizations, from (\ref{Eqn_1N}), we obtain:
\begin{equation}
\tau^m_r - \tau^b_r 
=  {\tau}_{sf}^m-  \tau_{sf}^b 
= {\tau}_{ss}^m-\tau_{sf}^b.
\label{Eqn_5N_1}
\end{equation}
{\bf IV. Fluid-type landslide and solid-type bed:} If the landslide is a fluid-type material and the bed is a solid-type material (flood entraining soil, sand, or gravel from the bed), then the fluid stress does not need to be distributed but only applied to the solid in the bed. 
 With these realizations, from (\ref{Eqn_2N}), we obtain:
\begin{equation}
\tau^m_r - \tau^b_r 
=  {\tau}_{fs}^m-  \tau_{fs}^b 
= {\tau}_{fs}^m-\tau_{ss}^b.
\label{Eqn_5N_2}
\end{equation}
{\bf V. Mixture landslide and solid-type bed:} If the landslide is a mixture material and the bed is a solid-type material (debris flow entraining soil, sand, or gravel from the bed), then landslide shear stresses do not need to be distributed, and only the bed solid shear resistance should be distributed.  With these realizations, from (\ref{Eqn_1N}) and (\ref{Eqn_2N}),
we get: 
\begin{eqnarray}
\begin{array}{lll}
\tau^m_r - \tau^b_r \!\!\!&=&\!\!\!\left [ \tau_{ss}^m - \alpha_f^b\tau_{ss}^b \right ] + 
                 \left [ \tau_{fs}^m - \alpha_s^b\tau_{fs}^b \right ]
                \!=\!\left [ \tau_{ss}^m - \alpha_f^b\tau_{ss}^b \right ] + 
                 \left [ \tau_{fs}^m - \alpha_s^b\tau_{ss}^b \right ]\\[3mm]
                \!\!\!&=&\!\!\!\left [ \tau_{ss}^m + \tau_{fs}^m \right ]- \tau_{ss}^b
                \!=\!\left [ \tau_{ss}^m - \tau_{ss}^b \right ] + \tau_{fs}^m.
\label{Eqn_6N}
\end{array}
\end{eqnarray}
If there is no fluid-solid interactions, i.e., only solid-type landslide and solid-type bed, this can easily be understood as it directly reduces to 
 $\left [ \tau_{ss}^m - \tau_{ss}^b \right ] = \left [ \tau_{s}^m - \tau_{s}^b \right ]$ which is (\ref{Eqn_4N}). If there is no solid but only fluid-type material in the landslide, then (\ref{Eqn_6N}) rather reduces to 
 $ \left [\tau_{fs}^m - \tau_{ss}^b \right ]$, which is (\ref{Eqn_5N_2}).
\\[3mm]
{\bf VI. Mixture landslide and fluid-type bed:} If the landslide is a mixture material and the bed is a fluid-type material (debris flow entraining river, lake or reservoir fluid), then, from (\ref{Eqn_1N}) and (\ref{Eqn_2N}), we have
\begin{equation}
\tau^m_r - \tau^b_r 
            = \left [ \tau_{sf}^m - \alpha_s^b\tau_{sf}^b \right ] +
              \left [ {\tau}_{ff}^m - \alpha_f^b\tau_{ff}^b \right ]
            = \left [ \tau_{ss}^m - \alpha_s^b\tau_{sf}^b \right ] +
            \left [ {\tau}_{ff}^m - \alpha_f^b\tau_{ff}^b \right ]
            = \left [ \tau_{ss}^m  + {\tau}_{ff}^m\right ] - \left [ \alpha_s^b\tau_{sf}^b + \alpha_f^b\tau_{ff}^b\right ].
\label{Eqn_7N}
\end{equation}
Unlike the same type of solid resistances from the bed in {\bf V}, here the fluid resistances from the bed $\tau_{sf}^b$ and $\tau_{ff}^b$ are incomparably different, as have been revealed in (\ref{Eqn_2NN}) and (\ref{Eqn_3NN}). So, the situation here could not be simplified further. This is an important novel development. However, if there is no solid in the landslide, then $\alpha_f^b = 1$ and (\ref{Eqn_7N})  reduces to 
$\left [ \tau_{ff}^m  - \tau_{ff}^b\right ] = \left [ \tau_{f}^m  - \tau_{f}^b\right ]$ which is (\ref{Eqn_5N}). In another scenario, if the landslide only contains the solid material, then $\alpha_s^b = 1$ and (\ref{Eqn_7N}) takes the form 
$\left [\tau_{ss}^m  - \tau_{sf}^b\right ]$, which is (\ref{Eqn_5N_1}).
\\[3mm]
{\bf VII. Solid-type landslide and mixture bed:} If the landslide is dry and bed material is a mixture (dry landslide of soil, sand, or gravel entraining debris material from the slope or bed), then, from (\ref{Eqn_1N}), we have:
\begin{equation}
\tau^m_r - \tau^b_r 
= \left [ \alpha_f^b\tau_{ss}^m - \tau_{ss}^b \right ] + 
  \left [ \alpha_s^b\tau_{sf}^m - \tau_{sf}^b \right ]
= \left [ \alpha_f^b\tau_{ss}^m - \tau_{ss}^b \right ] + 
             \left [ \alpha_s^b\tau_{ss}^m - \tau_{sf}^b \right ]
= \tau_{ss}^m -  \left [\tau_{ss}^b +  \tau_{sf}^b\right ]
= \left [\tau_{ss}^m  -  \tau_{ss}^b\right ] - \tau_{sf}^b. 
\label{Eqn_8N}
\end{equation}
Again, if there is no solid-fluid interactions (no fluid in the bed), this reduces to 
$\left [\tau_{ss}^m  -  \tau_{ss}^b\right ] = \left [\tau_{s}^m  -  \tau_{s}^b\right ]$ 
which is (\ref{Eqn_4N}).
In another scenario, if the bed only contains fluid material, then (\ref{Eqn_8N}) takes the form 
$\left [\tau_{ss}^m - \tau_{sf}^b\right ] $, which is (\ref{Eqn_5N_1}).
\\[3mm]
{\bf VIII. Fluid-type landslide and mixture bed:} If the landslide is a fluid material and bed is a mixture (flood entraining debris material from the slope or bed), then, from (\ref{Eqn_2N}), we have:
\begin{eqnarray}
\begin{array}{lll}
\tau^m_r - \tau^b_r\!\
=\!\left [ \alpha_s^b\tau_{fs}^m - \tau_{fs}^b \right ] \!+\! 
             \left [ \alpha_f^b {\tau}_{ff}^m - \tau_{ff}^b \right ]
\!=\!\left [\alpha_s^b \tau_{fs}^m  + \alpha_f^b {\tau}_{ff}^m \right ]
\!-\! \left [\tau_{fs}^b + \tau_{ff}^b \right ]
\!=\! \left [\alpha_s^b \tau_{fs}^m  + \alpha_f^b {\tau}_{ff}^m \right ]
\!-\! \left [\tau_{ss}^b + \tau_{ff}^b \right ].
\label{Eqn_9N}
\end{array}
\end{eqnarray}
Because of the complex fluid shear stresses, the stress jump mechanism here is quite different than that in {\bf VII}, and these shear stresses can not be reduced any further. However, if there is no solid in the bed, then  $\alpha_s^b = 0$, $\alpha_f^b = 1$ and (\ref{Eqn_9N}) reduces to $\left [\tau_{ff}^m - \tau_{ff}^b \right ]$ = $\left [\tau_{f}^m - \tau_{f}^b \right ]$, which is (\ref{Eqn_5N}). 
In another scenario, if the bed only contains the solid material, then $\alpha_s^b = 1$ and (\ref{Eqn_9N}) takes the form 
$\left [\tau_{fs}^m - \tau_{ss}^b\right ] $, which is (\ref{Eqn_5N_2}).
\\[3mm]
The reduced net shear stresses {\bf I}$-${\bf VIII} clearly demonstrate the consistency and physical significance of the interfacial interactions (\ref{Eqn_1N}) and (\ref{Eqn_2N}) and the shear stress jumps in (\ref{Eqn_3N}). These aspects have been elaborated at Section 6.1.
}

\subsection{Modelling basal substrate as an effectively single-phase material}

For simplicity and convenience, in applications, alternatively, we can also consider the basal substrate as an effectively single-phase (mixture) in which the dynamics of the fluid component is not explicit, but only via the pore pressure of the fluid in the solid matrix. In this situation, commonly, a single solid-type frictional (for simplicity, ignoring the collisional stress that can be easily added if needed) shear stress (the total shear stress) is considered for the erodible bed as:
\begin{equation}
\tau^b = \left ( 1 - \frac{p_f^b}{p_T^b}\right ) \rho^b_B g^z h\mu^b_s,
\label{Eqn_4_1}
\end{equation}
 where, $p_f^b$ and $p_T^b$ are the pore fluid pressure and the total pressure in
the basal material and ${p_f^b}/{p_T^b}$ represents the corresponding pore pressure ratio (ratio between the basal pore fluid pressure and the total basal normal stress, Hungr, 1995; Pudasaini, 2012), and $\rho^b_B$ is the bulk density of the basal substrate.
$p_T^b$ is given by $p_T^b = \alpha_s^b\rho_s^b g^z h + \alpha_f^b\rho_f^b g^z h$. However, depending on the dynamical state of the pore fluid pressure, $p_f^b$ varies between $\alpha_f^b\rho_f^b g^z h$ and $\alpha_s^b\rho_s^b g^z h + \alpha_f^b\rho_f^b g^z h$. So, there exists a parameter (or a function) $\Upsilon^b \in [0, 1]$ such that 
\begin{equation}
p_f^b = \left ( 1- \Upsilon^b\right ) \alpha_f^b\rho_f^b g^z h+ \Upsilon^b \left ( \alpha_s^b\rho_s^b g^z h+ \alpha_f^b\rho_f^b g^z h\right). 
\label{Eqn_4p}
\end{equation}
We call $\Upsilon^b$ the effective pore pressure ratio.
The use of this terminology is explained below.
This expression automatically satisfies the natural condition that as $\Upsilon^b \to 0$ the fluid pressure nears the usual hydrostatic pressure, and as $\Upsilon^b \to 1$ the fluid pressure approaches the total material load (pressure). As $\Upsilon^b$ covers the whole spectrum, 
it measures the deviation of the pore fluid pressure from the hydrostatic fluid pressure to the complete liquefaction of the bed material. With a simple algebra, (\ref{Eqn_4_1}) can be re-written as:
\begin{equation}
\tau^b = \left ( 1 - \Upsilon^b\right) {\mathcal S_c} \,\rho^b_B g^z h\mu^b,
\label{Eqn_4}
\end{equation}
where 
\begin{equation}
{\mathcal S_c} = \left[\frac{\rho_s^b\alpha_s^b}{\rho_s^b\alpha_s^b + \rho_f^b\alpha_f^b}\right],
\label{Eqn_4_0}
\end{equation}
and ${\mathcal S_c} \in [0, 1]$. We call ${\mathcal S_c}$ the shear stress correction factor. 
As revealed below, the representation of $\tau^b$ in (\ref{Eqn_4}) has several crucial mechanical inferences. 
\\[3mm]
($i$) When there is a substantial amount of fluid in the erodible bed substrate, simply using $ 1 - \Upsilon^b$ and not considering ${\mathcal S_c}$ significantly overestimates the basal shear stress. This indicates that, the previous (frictional) bulk basal shear stress models (e.g., Hungr, 1995; Pudasaini et al., 2005; 
Iverson and  George, 2014; Iverson and Ouyang, 2015) 
for the mixture material are inappropriate. So, in fact, the basal material is substantially weaker than it was previously thought. This is a fundamentally novel understanding with important implications in mass flow simulation, particularly when it comes to the erosive flows. 
\begin{figure}
\begin{center}
\includegraphics[width=9cm]{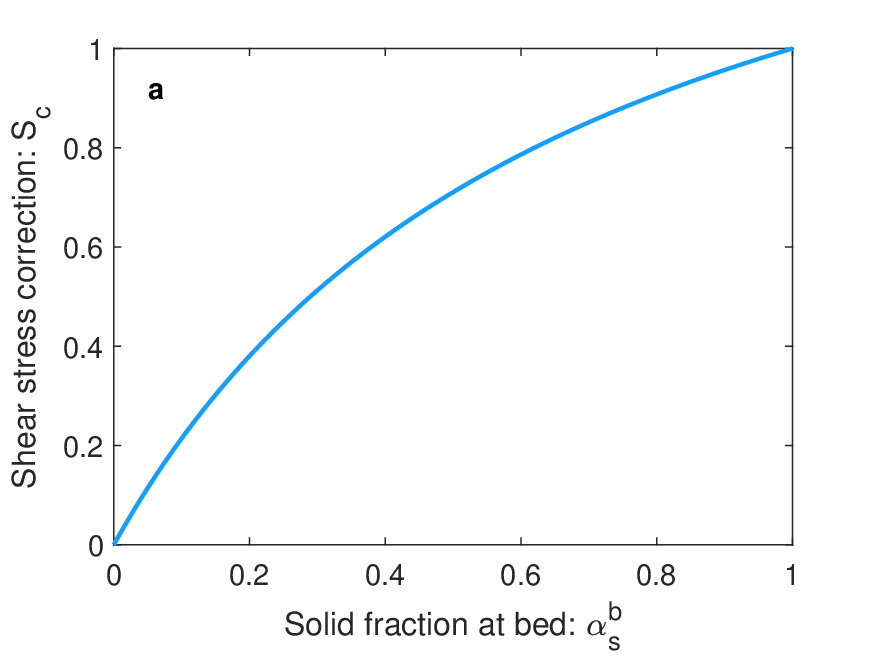}
\includegraphics[width=9cm]{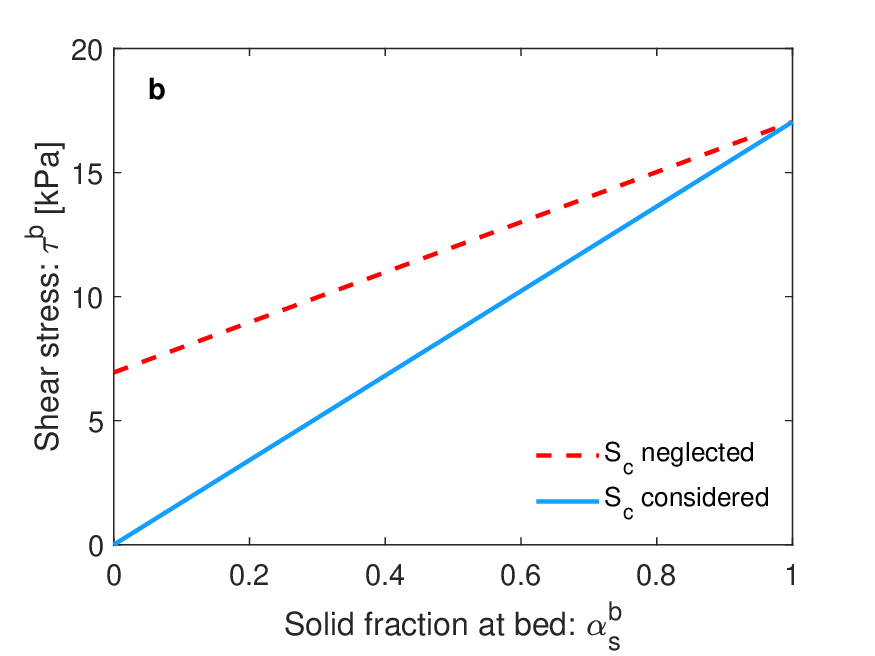}
  \end{center}
  \caption[]{(a) Shear stress correction factor $\mathcal S_c$ as given by (\ref{Eqn_4_0}) that increases with the solid volume fraction $\alpha_s^b$ in the bed material. (b) Incorrect and correct shear stresses that respectively neglect and include $\mathcal S_c$. The incorrect shear stress substantially overestimates the basal shear resistance, largely so for the lower $\alpha_s^b$.}
  \label{Fig_1}
\end{figure}
\\[3mm]
($ii$) Another crucial aspect is the direct involvement of the solid particle concentration in the basal shear stress via $\mathcal S_c$, which was ignored by all the previous shear stress models as those models only use the mixture bulk density $\rho^b_B$, but not the composite mixture density $\rho^b$. 
From the mixture perspective of the basal substrate, i.e., $\rho^b_B = \rho^b$, (\ref{Eqn_4}) reduces to 
\begin{equation}
\tau^b = \left ( 1 - \Upsilon^b\right) \alpha_s^b\rho_s^b \,g^z h\mu^b_s.
\label{Eqn_4_4}
\end{equation}
Mechanically, the form of (\ref{Eqn_4_4}) is important.
Since the particle concentration is one of the main controlling dynamical factors in determining the shear resistance, those models (e.g., Hungr, 1995; Pudasaini et al., 2005; Iverson and  George, 2014; Iverson and Ouyang, 2015) without its involvement cannot legitimately represent the true shear resistance. 
\\[3mm]
To see the dynamical effect of the factor ${\mathcal S_c}$ in the basal shear resistance $\tau^b$, consider some usual parameters (Mergili et al., 2020, Pudasaini and Fischer, 2020a; Pudasaini and Krautblatter, 2021): $\rho_s^b = 2700, \rho_f^b = 1100, h = 5, \delta_s^b = 20^\circ, \zeta = 45^\circ, g = 9.81$, where $\mu_s^b = \tan\delta_s^b$ is the friction coefficient, $\zeta$ is the slope angle, $g^z = g \cos\zeta$, and $g$ is the gravitational acceleration. Furthermore, we consider $\Upsilon^b = 0.5$.
As shown in Fig. \ref{Fig_1}a, the expression in (\ref{Eqn_4}) reveals the fact that the shear resistance decreases non-linearly with the decreasing solid fraction in ${\mathcal S_c}$, it tends to vanish as the solid fraction is negligibly small, and it takes the maximum value as the solid fraction approaches unity. Only this limiting value could be considered by all the existing effectively single-phase shear resistance models (e.g., Hungr, 1995; Pudasaini et al., 2005; Iverson and  George, 2014; Iverson and Ouyang, 2015), which is less realistic for the mixture substrate. 
For a particularly representative value of the basal solid fraction $\alpha_s^b = 0.55$, $\mathcal S_c = 0.75$. 
This means, the existing shear resistance models typically overestimates 
$\tau^b$ by about 25\%, which is substantial. The overestimation can further increase if the basal material is till-reach for which the true solid density can decrease to $\rho_s^b = 2000$ and the shear stress reduces by about 30\%. This can result in completely different scenario as (erosive) debris avalanches travel long distances. This demonstrates the mechanical and dynamical importance of the new shear resistance model in (\ref{Eqn_4}) and (\ref{Eqn_4_4}). 
Similarly, Fig. \ref{Fig_1}b shows that the classical shear stress that neglects ${\mathcal S_c}$ substantially overestimates the basal shear resistance, mainly for the lower $\alpha_s^b$.
\\[3mm]
$(iii)$ The shear stress given by (\ref{Eqn_4}) or (\ref{Eqn_4_4}) includes both the aspects of pore pressure ratio and its dependency on the particle concentration. We have presented the first shear stress model that formally includes these important mechanical aspects. 
\\[3mm]
{\bf The effective pore pressure ratio:} 
 Following its definition from (\ref{Eqn_4p}), $\Upsilon^b$ can be explicitly written as:
\begin{equation}
\Upsilon^b = \frac{p_f^b - \alpha_f^b\rho_f^b g^z h}{\alpha_s^b\rho_s^b g^z h},
\label{Eqn_4_33}
\end{equation}
which says that $\Upsilon^b$ is the measure of the actual pore fluid pressure in excess to the hydrostatic fluid pressure (relative to the solid load in the bed). The structure (form) of $\Upsilon^b$ in (\ref{Eqn_4_33}) justifies the terminology used above for $\Upsilon^b$, the effective pore pressure ratio.  

\section{Erosion rates}

\subsection{The total erosion rate}

Erosion rate (or the rate of mass production) is determined with the jump in the shear stresses and the jump in the momentum fluxes across the landslide-bed interface.
We consider the jump in the shear stresses, $\tau^m - \tau^b $ from (\ref{Eqn_3N}) together with all the stress components as modelled in (\ref{Eqn_1NN})-(\ref{Eqn_3NN}), and later ((\ref{Eqn_Col1})-(\ref{Eqn_Col11_a})) (alternatively, (\ref{Eqn_6N}) with (\ref{Eqn_4}) or (\ref{Eqn_4_4}), $\left [\tau^m_{ss} + \tau^m_{fs}\right] - \tau^b_{ss}$, $\tau^b = \tau^b_{ss}$), 
and the jump
in the momentum fluxes, $\rho^m u_l^m - \rho^b u^b$ across the erosion-interface.
Then, following Pudasaini and Fischer (2020a), the total erosion-rate $E$ of the system is
obtained by:
\begin{equation}
E = \frac{\tau^m - \tau^b}{ \rho^m u_l^m - \rho^b u^b }
=  \frac{\tau^m - \tau^b}{\left ( \rho^m\lambda^m_l - \rho^b\lambda^b \right
) u^m },
\label{Eqn_6}
\end{equation}
where, the mixture densities $\rho^m, \rho^b$ are given by (\ref{Eqn_1}) and (\ref{Eqn_2}), respectively,
and the drift relations for $u_l^m$ and $u^b$ have been employed from (\ref{Eqn_5}) with their descriptions at Section 3.1.3. We note that negative of $E$
 corresponds to the deposition. 
 \\[3mm]
Now, following Pudasaini and Fischer (2020a), we consider the shear velocity $u^*$ of the system which is given by the square root of the ratio between the net shear stress of the system and the relevant net density across the erosion-interface, and reads:  
\begin{equation}
\displaystyle{u^* = \frac{\sqrt{\tau^m - \tau^b }}
 {\sqrt{\left ( \rho^m\lambda^m_l - \rho^b\lambda^b \right )}}}. 
 \label{Eqn_6_sv}
\end{equation}
The shear velocity is proportional to the flow velocity $u^m$. So, with the proportionality factor
$\tilde \nu$, we can define a
relationship as $u^m = \tilde \nu u^*$.
With this, the erosion rate $E$ yields
\begin{equation}
E = \frac{\sqrt{\tau^m - \tau^b }} {\sqrt{\nu \left ( \rho^m\lambda^m_l -
\rho^b\lambda^b \right )}},
  \label{Eqn_7}
\end{equation}
where, for simplicity, $\nu = \tilde \nu^2$ is set. 
 The total (overall) erosion rate $E$ in (\ref{Eqn_7}) is, in fact, in a very compact form. While expressing the involved total shear stresses $\tau^m$, $\tau^b$, and the mixture densities $\rho^m, \rho^b$ in explicit form, this expression appears to be extensive. 
 The number of involved model (physical) parameters, their values and closures are presented at Section 6.4.

\subsubsection{Erosive-shear-velocity}

By comparing (\ref{Eqn_6}) with (\ref{Eqn_6_sv}), we obtain an expression for $u^*$. This leads to the definition of a new shear velocity, which we call the erosive-shear-velocity, write as $u_{_{{\mathcal S}_E}}$, and takes the form:
\begin{equation}
\displaystyle{u_{_{{\mathcal S}_E}} 
= \mathcal P_{_{E_u}}\, u^m},
 \label{Eqn_6_sv1}
\end{equation}
where, $P_{_{E_u}} = \sqrt{{E}/{u^m}}$ is the dynamical proportionality factor 
between the flow velocity and the erosive-shear-velocity.
The erosive-shear-velocity $u_{_{{\mathcal S}_E}}$ is the square root of the momentum production induced by the flow velocity and the erosion rate $\left (\sqrt{u^m E}\right)$. 
$u_{_{{\mathcal S}_E}}$ increases with the flow velocity $u^m$ (but can be directly related to the erosion velocity $u^b$ with the relationship $u^b = \lambda^b u^m$) and the erosion rate $E$. So, the erosive-shear-velocity is primarily induced by erosion, and vanishes for non-erosive flows. It is important to note that the erosive-shear-velocity is proportional to the flow velocity, and its proportionality $P_{_{E_u}}$ varies as a function of the erosion rate and inversely with the flow velocity, i.e., $\sqrt{{E}/{u^m}}$. As the erosion rate can be relatively small and the flow velocity can be relatively large, the erosive-shear-velocity is expected to be substantially smaller than the shear velocity. Yet, as clear from their definitions, the erosive-shear-velocity is fundamentally different than the classical shear velocity as it includes the erosion rate.
The novel mechanism and understanding of erosive-shear-velocity is revealed here with our dynamical modelling approach. 

\subsubsection{Closure for the shear velocity} 

Usually, the shear velocity is about
5\% to 10\% of the mean flow velocity, and thus, $1/\sqrt{\nu}\in \left (0.05, 0.1 \right)$. So, for simplicity, we can accordingly take the
suitable value of $\nu$ in the range (100, 400). Otherwise, we follow
Pudasaini and Fischer (2020a) for an analytical closure relation for $\nu$.

\subsubsection{Closures for erosion drifts}

\begin{figure}
\begin{center}
\includegraphics[width=9cm]{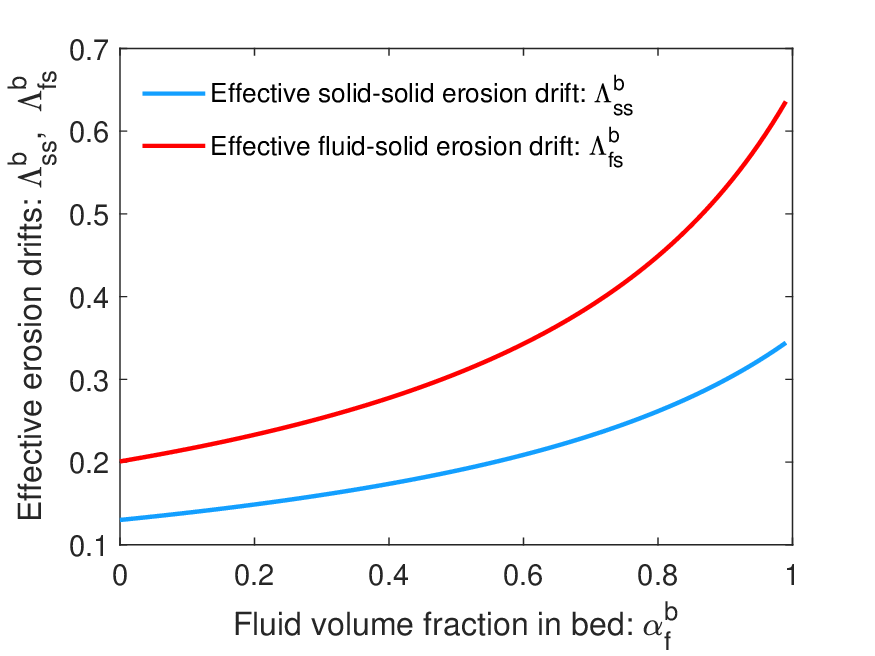}
\includegraphics[width=9cm]{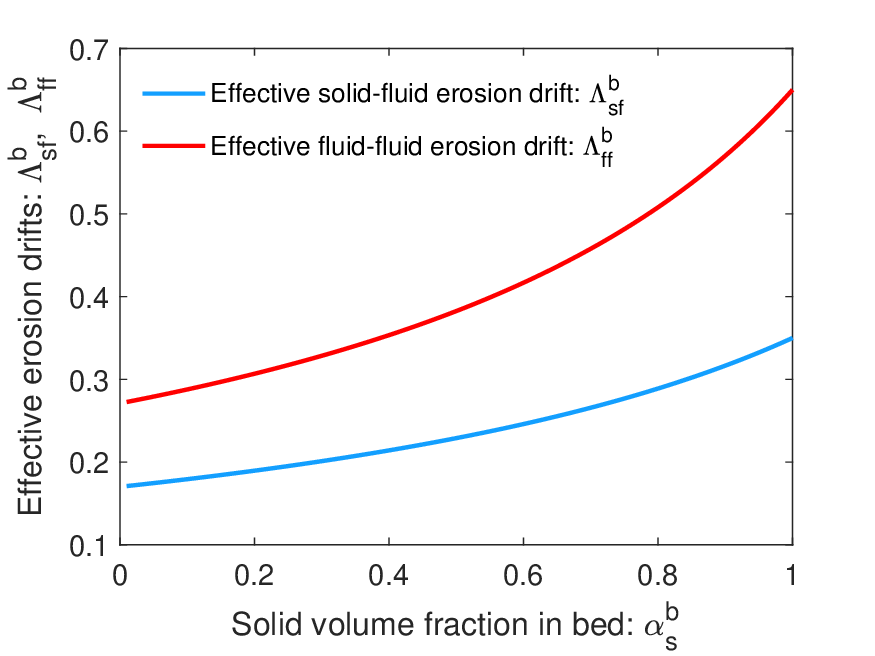}
  \end{center}
  \caption[]{Left: the solid-solid ($ss$) and fluid-solid ($fs$) effective erosion drifts associated with the erosion velocity of the solid particle from the bed. Right: the fluid-fluid ($ff$) and solid-fluid ($sf$) effective erosion drifts associated with the erosion velocity of the fluid from the bed.}
  \label{Fig_4}
\end{figure}

Erosion drifts are essential quantities as they provide crucial information about the erosion velocities which play central role in explaining the erosion rate and the net momentum production, and the associated excess energy that control the mobility of erosive mass transports (Pudasaini and Krautblatter, 2021). So, now we construct different erosion drift equations providing mechanical closures for all the erosion drifts appearing in the process of model development (erosion rates and net momentum productions). 
\\[3mm]
{\bf I. Total erosion drift, solid-solid and fluid-fluid direct phase erosion drifts:} Considering the balance between the effective reduced net (frictional plus collisional)
stress, $\left ( \tau^m - \tau^b\right )/\rho^m$, where $\rho^m$ emerges due to the mass factor, and momentum production in erosion, which is related to (\ref{Eqn_6}), $u^b E = \lambda^b u^m E = \lambda^b \left ( \tau^m - \tau^b\right )/\left ( \rho^m\lambda^m_l - \rho^b\lambda^b\right )$ (Pudasaini and Fischer, 2020a), we obtain
the erosion drift equation for the total mixture system as:
\begin{equation}
\lambda^m_l = \left ( 1 + \frac{\rho^b}{\rho^m}\right)\lambda^b.
\label{Eqn_8}
\end{equation}
This is a compact, strong and general drift equation, where $N^I = {\rho^b}/{\rho^m}$ is the associated (erosional) bed inertial number as the ratio between the effective mass in the bed and the flow (Pudasaini and Krautblatter, 2021).
\begin{figure}
\begin{center}
\includegraphics[width=9cm]{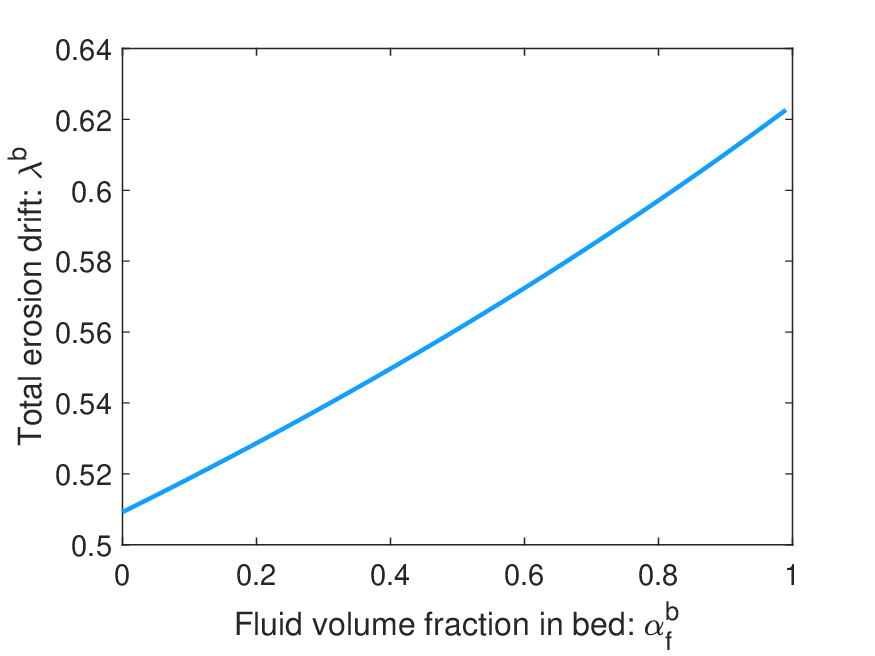}
  \end{center}
  \caption[]{The total erosion drift $\lambda^b$ given by (\ref{Eqn_8}) as a function of the basal fluid volume fraction $\alpha_f^b$ showing the increase in $\lambda^b$ with the weakened basal material.}
  \label{Fig_2}
\end{figure}
We analyze each of the erosion drift (for simplicity) by locally keeping the associated mass of the landslide unchanged but weakening the basal materials. Otherwise, we can also reverse the situation, or consider general settings and analyze the erosion drifts.
As we will see later, one of the main aspects is to demonstrate significance to dominant influence of the cross-erosion drifts in the erosion dynamics, we consider the fluid dominated landslide, akin to hyperconcentrated flows. Without loss of generality, this is achieved by choosing $\alpha_s^m = 0.35$ (so $\alpha_f^m = 0.65$).
Moreover, following the literature (Mergili et al., 2020; Pudasaini and Krautblatter, 2021) other physical parameters used are:
$\rho_s^m = 2700, \rho_f^m = 1100; \rho_s^b = 1600, \rho_f^b = 1000; 
%\alpha_s^m = 0.35,\alpha_f^m = 1 - \alpha_s^m; 
u_s^m = 25, u_f^m = 30$.
With these, $\rho_s^m\alpha_s^m$ and $\rho_f^m\alpha_f^m$ are assumed to be locally not changing so much, which however, are variable in general situations.
%=  2700*0.35 = 945; 1100*0.65 = 715 can be assumed to be locally not changing so much.
\\[3mm]
Figure \ref{Fig_2} displays the dynamics of the total erosion drift $\lambda^b$ as a function of the basal fluid volume fraction $\alpha_f^b$. As $\alpha_f^b$ increases, the bed material becomes weaker. It results in the increased values of the total erosion drift, which in turn leads to an increase in erosion velocity. This is how the drift explains the erosion mechanism.
There are two important features associated with the drift equation (\ref{Eqn_8}).  
\\[3mm]
{\bf A. Solid-solid and fluid-fluid direct-phase-erosion drifts:} In the limit, (\ref{Eqn_8}) reduces to the solid
only (solid-solid) and fluid only (fluid-fluid) erosion drift equations as constructed by Pudasaini and Fischer (2020a):
\begin{equation}
\lambda_{s_l}^m = \left ( 1 +
\frac{\rho_s^b\alpha_s^b}{\rho_s^m\alpha_s^m}\right)\lambda_{ss}^b,
\label{Eqn_9}
\end{equation}
and
\begin{equation}
\lambda_{f_l}^m = \left ( 1 +
\frac{\rho_f^b\alpha_f^b}{\rho_f^m\alpha_f^m}\right)\lambda_{ff}^b,
\label{Eqn_10}
\end{equation}
respectively. In these drift equations $N_{ss}^I = {\rho_s^b\alpha_s^b}/{\rho_s^m\alpha_s^m}$, and 
$N_{ff}^I = {\rho_f^b\alpha_f^b}/{\rho_f^m\alpha_f^m}$ are the corresponding bed inertial numbers.
\\[3mm]
{\bf B. Clousers:} As the velocity shearing in the depth averaged modelling frame is
generally ignored, $\lambda^m_l$ in (\ref{Eqn_8}) can be set to unity (Pudasaini and Fischer, 2020a). So, since $\rho^m$ and
$\rho^b$ are known from the flow configuration, the drift equation (\ref{Eqn_8})
provides a closer for 
$\lambda^b$. This also applies to (\ref{Eqn_9}) and (\ref{Eqn_10}), providing the closures for $\lambda_{ss}^b$ and $\lambda_{ff}^b$. 
\\[3mm]
{\bf II. Solid-fluid and fluid-solid cross-erosion-drifts:}
We also need to develop closure relations for the cross-drifts $\lambda_{sf}^b$ and $\lambda_{fs}^b$ whose structures are not known yet. With the elegant procedure, first, we construct the closure for $\lambda_{sf}^b$. For this, consider the fluid erosion rate $E_f$. It can be formally de-composed into the components induced by the fluid-fluid $\left (E_{ff}\right)$ and the solid-fluid $\left (E_{sf}\right)$ interactions between the landslide and the bed. 
As the cross erosion drift $\lambda_{sf}^b$ is associated with the solid-fluid interaction, we need to consider the cross-contribution $E_{sf}$. 
Following the structure of the erosion rate
from Section 3.1, we can write $E_{sf}$ that is associated with its corresponding jump in the shear stresses $\alpha_s^b \left [ \tau_{sf}^m - \tau_{sf}^b\right ]$, and the jump in the momentum fluxes $\left [ \alpha_s^m\rho_s^m u_{{sf}_l}^m - \alpha_f^b\rho_f^b u_{{sf}}^b\right ]$, and applying the relevant drifts for $u_{{sf}_l}^m$ and $u_{{sf}}^b$ from (\ref{Eqn_5}), as
\begin{equation}
\displaystyle{E_{sf} 
= \frac{\alpha_s^b \left [ \tau_{sf}^m - \tau_{sf}^b\right ]}{\left [ \alpha_s^m\rho_s^m u_{{sf}_l}^m - \alpha_f^b\rho_f^b u_{{sf}}^b\right ]}
= \frac{\alpha_s^b \left [ \tau_{sf}^m - \tau_{sf}^b\right ]}{\left [ \alpha_s^m\rho_s^m \lambda_{{sf}_l}^m - \alpha_f^b\rho_f^b \lambda_{{sf}}^b\right ]u_s^m}.
}
\label{Eqn_10_sf_1}
\end{equation}
So, since $u_{sf}^b = \lambda_{sf}^b u_s^m$, the corresponding momentum production $\left ( u_{sf}^bE_{sf} \right )$ takes the form:
\begin{equation}
\displaystyle{u_{sf}^bE_{sf} 
= \frac{\alpha_s^b \left [ \tau_{sf}^m - \tau_{sf}^b\right ] \lambda_{sf}^b}{\left [ \alpha_s^m\rho_s^m \lambda_{{sf}_l}^m - \alpha_f^b\rho_f^b \lambda_{{sf}}^b\right ]}.
}
\label{Eqn_10_sf_2}
\end{equation}
Moreover, the effectively reduced net (frictional plus collisional) stress (dissipation) for the solid-fluid interaction is given by,
\begin{equation}
\displaystyle{\frac{\alpha_s^b \left [ \tau_{sf}^m - \tau_{sf}^b\right ]}{\alpha_s^m\rho_s^m}.
}
\label{Eqn_10_sf_3}
\end{equation}
Since (\ref{Eqn_10_sf_2}) and (\ref{Eqn_10_sf_3}) are equivalent (Pudasaini and Fischer, 2020a), by comparing these expressions, we finally obtain a closure relationship between $\lambda_{{sf}_l}^m$ and $\lambda_{{sf}}^b$ as
\begin{equation}
\displaystyle{\lambda_{{sf}_l}^m
= \left ( 1 + \frac{\rho_f^b\alpha_f^b}{\rho_s^m\alpha_s^m}\right )\lambda_{{sf}}^b,
}
\label{Eqn_10_sf}
\end{equation}
in which $N_{sf}^I = {\rho_f^b\alpha_f^b}/{\rho_s^m\alpha_s^m}$ is the associated bed inertial number.
Although (\ref{Eqn_10_sf}) involved a bit of mechanical derivation (as seen above), physically this is a great achievement. The point is that, once derived formally and known explicitly, this can also be directly extracted now from the compact drift equation (\ref{Eqn_8}) by carefully and consistently taking the cross terms $\rho_f^b\alpha_f^b$ from $\rho^b$ and $\rho_s^m\alpha_s^m$  from $\rho^m$, and correspondingly realizing $\lambda^b$ as $\lambda_{{sf}}^b$ and $\lambda^m_l$ as $\lambda_{{sf}_l}^m$. This converts (\ref{Eqn_8}) into (\ref{Eqn_10_sf}). 
\\[3mm]
However, note that $E_{sf}$ as considered in (\ref{Eqn_10_sf_1}) is only for the purpose of constructing the closure relation (\ref{Eqn_10_sf}) but not for the purpose of obtaining a contribution for the fluid erosion rate $E_{f}$. Due to its consistency and importance in practical applications (as explained at Section 2.1 and Section 6.3), the solid and fluid erosion rates $E_{s}$ and $E_{f}$ are derived at Section 3.4 from the total erosion rate $E$ in (\ref{Eqn_7}).
\\[3mm]
Exactly in the same manner, we can construct the closure relationship between $\lambda_{{fs}_l}^m$ and $\lambda_{{fs}}^b$, either formally as in (\ref{Eqn_10_sf}), or by the careful and consistent extraction from (\ref{Eqn_8}) to yield:
\begin{equation}
\displaystyle{\lambda_{{fs}_l}^m
= \left ( 1 + \frac{\rho_s^b\alpha_s^b}{\rho_f^m\alpha_f^m}\right )\lambda_{{fs}}^b,
}
\label{Eqn_10_fs}
\end{equation}
where $N_{fs}^I = {\rho_s^b\alpha_s^b}/{\rho_f^m\alpha_f^m}$ is the respective bed inertial number.
With this, all the drifts in (\ref{Eqn_5}) are known fully and mechanically without any fit parameter. 
\begin{figure}
\begin{center}
\includegraphics[width=9cm]{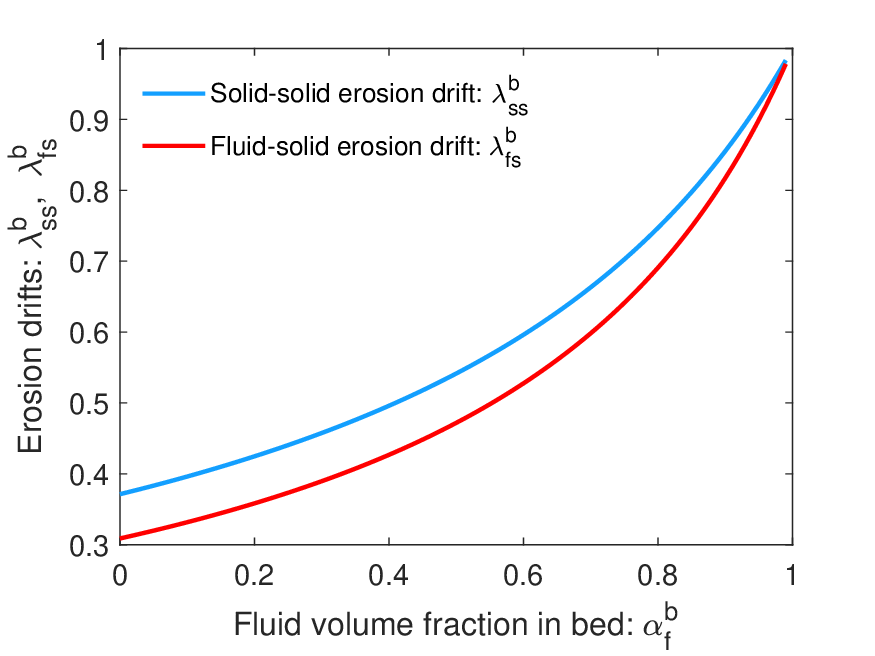}
\includegraphics[width=9cm]{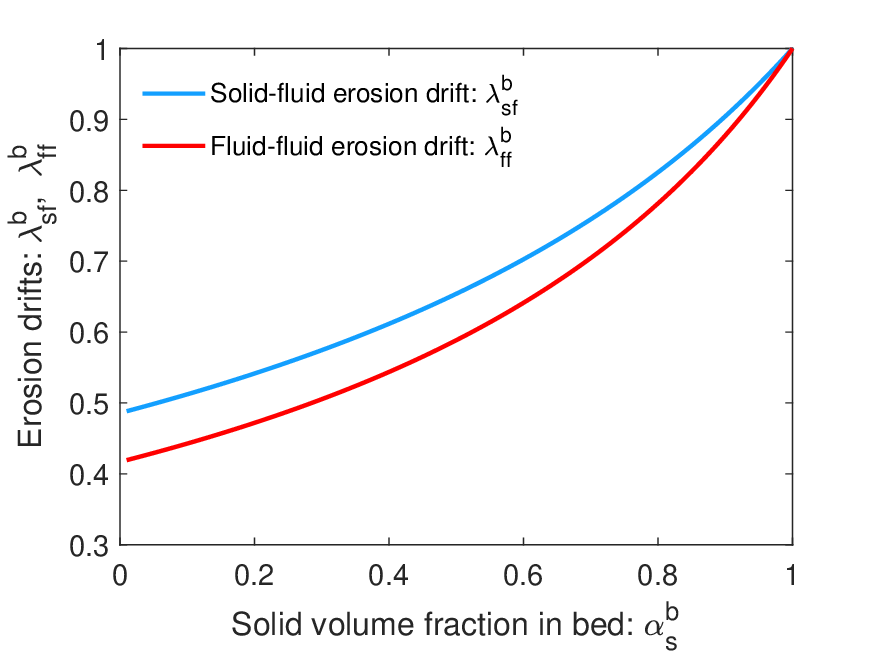}
  \end{center}
  \caption[]{Left: the solid-solid ($ss$) and fluid-solid ($fs$) erosion drifts associated with the erosion velocity of the solid particle given by (\ref{Eqn_9}) and (\ref{Eqn_10_fs}) in the bed. Right: the fluid-fluid ($ff$) and solid-fluid ($sf$) erosion drifts associated with the erosion velocity of the fluid given by (\ref{Eqn_10}) and (\ref{Eqn_10_sf}) in the bed.}
  \label{Fig_3}
\end{figure}
\\[3mm]
It is crucial to understand the dynamics of the phase-phase and cross-phase drifts in accordance of how they are influenced by changing composition of the bed material which controls the erosion velocities.   
Figure \ref{Fig_3} shows the phase-phase ($ss$ and $ff$) and cross-phase ($fs$ and $sf$) erosion drifts $\left (\lambda_{ss}^b, \lambda_{ff}^b\right)$, and $\left (\lambda_{fs}^b, \lambda_{sf}^b\right)$, respectively. As the erosion velocity of the solid particle at the bed $u_s^b$ is a composite function of the solid-solid ($ss$) and fluid-solid ($fs$) erosion drifts they are put together. Similarly, as the erosion velocity of the fluid molecule at the bed $u_f^b$ is a composite function of the fluid-fluid ($ff$) and solid-fluid ($sf$) erosion drifts they are put together. However, these sets of erosion drifts are described as a function of the fluid volume fraction in the bed $\left (\alpha_f^b\right)$ and the solid volume fraction in the bed $\left (\alpha_s^b\right)$, respectively. This is done logically and legitimately. Because, as the fluid volume fraction in the bed increases, there are less solid particles in the bed. Then, both for the solid and the fluid in the landslide, it will be easier to mobilize the solid in the bed. This is why both of $\lambda_{ss}^b$ and $\lambda_{fs}^b$ increase with $\alpha_f^b$. This is quite natural. Similar analysis applies to the mobilization of the basal fluid by the solid and the fluid from the landslide. Figure \ref{Fig_3} clearly manifests that the cross-drifts $\lambda_{fs}^b$ and $\lambda_{sf}^b$ may even be of the same order of magnitude as the phase-drifts $\lambda_{ss}^b$ and $\lambda_{ff}^b$. These important aspects could not be considered previously.
\\[3mm]
The drifts and cross drifts $\lambda_{ss}^b$ and $\lambda_{fs}^b$ are employed to describe the solid erosion velocity $u_s^b$ in (\ref{Eqn_1NNN}). However, they appear together with $\alpha_s^m$ and $\alpha_f^m$. So, it is more logical to define the effective phase-drifts and effective cross-drifts as $\Lambda_{ss}^b = \alpha_s^m\lambda_{ss}^b$ and $\Lambda_{fs}^b = \alpha_f^m\lambda_{fs}^b$. The results are presented in Fig. \ref{Fig_4}. The advantages here are two fold. First, the figure proves that the effective cross-drift $\Lambda_{fs}^b$ may even strongly dominate the effective phase-drift $\Lambda_{ss}^b$ and the difference between them increases steadily. Second, their sum is bounded from above by unity. This is so nice as they together constitute the velocity of the mobilized solid particle from the bed. Analogous analysis applies between the effective cross-drift $\Lambda_{sf}^b$ and the phase-drift $\Lambda_{ff}^b$ associated with the mobilization of the fluid molecule from the bed. 
\\[3mm]
{\bf III. The super-erosion-drift equation:} It is so pleasant to observe some important aspects of the erosion drift equation derived above. First, now the cross-drifts (\ref{Eqn_10_sf}) and (\ref{Eqn_10_fs}) can be directly extracted from the compact total drift relationship (\ref{Eqn_8}). For this reason, we call (\ref{Eqn_8}) the super-erosion-drift-equation (or simply the S-drift). Yet, the most astonishing fact is that the S-drift contains all the necessary information for all the drift factors. It means, in fact, essentially all the needed five drift factors can be obtained by a single drift equation (\ref{Eqn_8}). Second, the cross-drift (\ref{Eqn_10_sf}) and (\ref{Eqn_10_fs}) are symmetrical about the solid-fluid and fluid-solid cross-phase interactions across the erosion interface. Third, these properties could be expected, but they also signify the strength of the total erosion drift equation (\ref{Eqn_8}), consistency of all the derivations for drifts, and also the mechanical equivalence between the erosion-induced momentum production and the reduced (frictional plus collisional) strength for the erosional mass flows. 

\subsection{Bed inertial numbers and mobility}

The shear resistances of the bed materials (in total or component-wise, direct or cross-phases) against the applied shear stresses from the landslide are described by the bed inertial numbers $N^I, N^I_{ss}, N^I_{ff}, N^I_{sf}, N^I_{fs}$ in (\ref{Eqn_8}), (\ref{Eqn_9}), (\ref{Eqn_10}), (\ref{Eqn_10_sf}), (\ref{Eqn_10_fs}), respectively. As these numbers decrease, bed materials become weaker against the applied shear. This elevates the values of corresponding drifts $\lambda^b, \lambda^b_{ss}, \lambda^b_{ff}, \lambda^b_{sf}, \lambda^b_{fs}$, resulting in the increased erosion velocity. This effectively means the mass flow mobility is associated with the bed inertial numbers.
Depending on the respective inertial number, each erosion drift has its own special dynamics.
However, the structures of the cross-inertial-numbers $N^I_{sf}$ and $N^I_{fs}$ tell that it is relatively difficult for the fluid in the landslide to mobilize the grain in the bed, but it is relatively easy for the grain in the landslide to mobilize the fluid in the bed. This is intuitively clear from the perspective of the strength of material, but revealed here with the mechanical closures for the cross-erosion drifts. 
In this respect, it is important to carefully analyze the erosion velocities 
in relation to the involved erosion drifts, but also with the volume fractions of the solid and fluid in the landslide and their respective velocities. We focus on this below.

\subsection{Importance of different components in erosion velocities}

\begin{figure}
\begin{center}
\includegraphics[width=9cm]{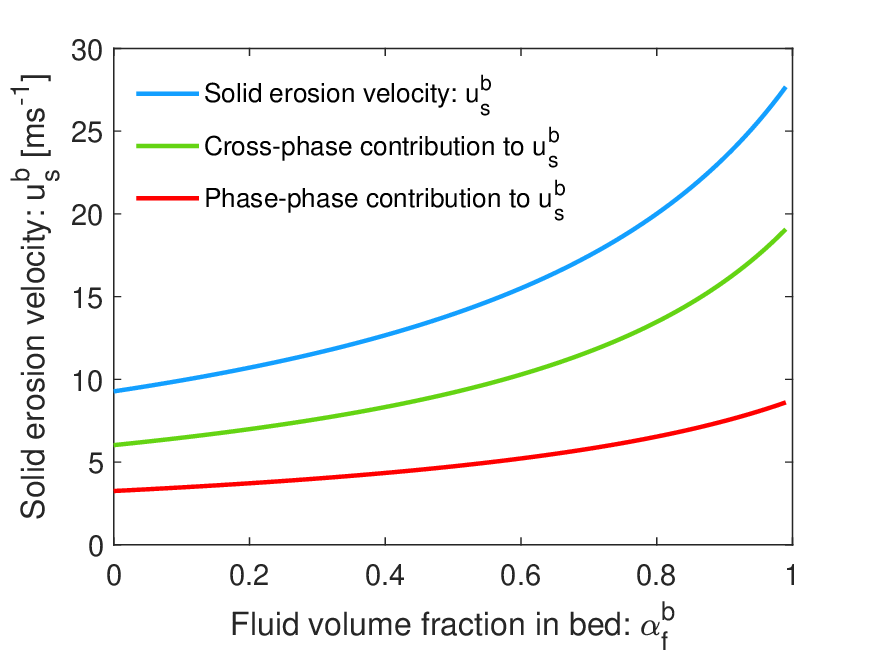}
\includegraphics[width=9cm]{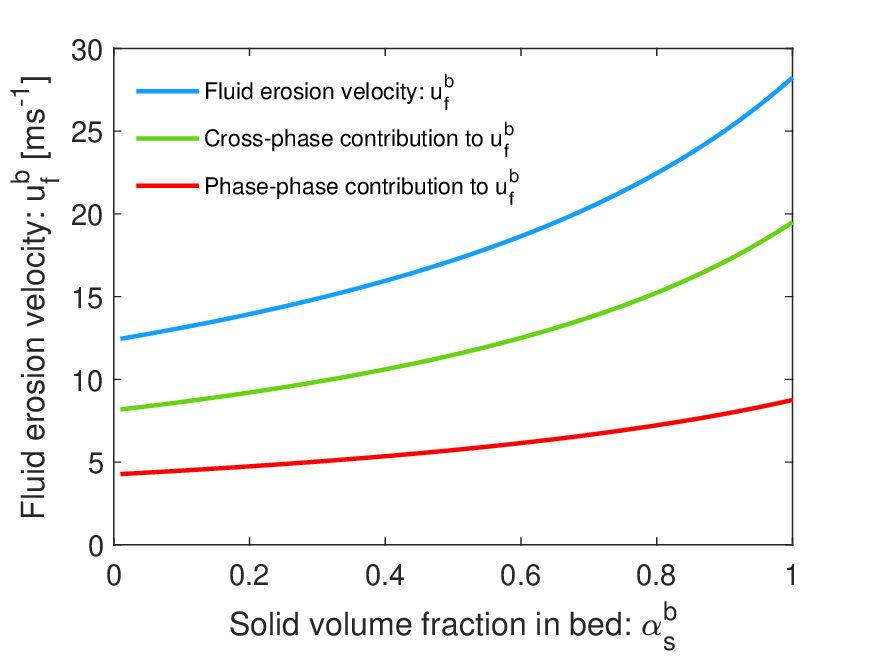}
  \end{center}
  \caption[]{Contributions of different components for the solid erosion velocity $u_s^b$ (left), and the fluid erosion velocity $u_f^b$ (right) in (\ref{Eqn_1NNN}) together with the phase- and cross-drifts and the solid and fluid volume fractions in the landslide.}
  \label{Fig_5}
\end{figure}

From (\ref{Eqn_1NNN}), by considering the solid erosion velocity $u_s^b = \alpha_s^m \lambda_{ss}^b u_s^m + \alpha_f^m \lambda_{fs}^b u_f^m$, we see that there are  complex and composite contributions to the erosion velocity from the fractions of the materials with their velocities, and the drifts. Usually, $\lambda_{ss}^b$ can be substantially greater than $\lambda_{fs}^b$. However, depending on the volume fractions and velocities, the component $\alpha_f^m \lambda_{fs}^b u_f^m$ may still play important to dominant role in comparison to the other component $\alpha_s^m \lambda_{ss}^b u_s^m$ in $u_s^b$. Example includes the mobilization of light and loose material (e.g., till) from the bed by fluid dominated debris flood with higher fluid velocity. So, the contribution of the bed solid mobilization by the fluid in the landslide cannot just be ignored, but was disregarded previously. We have shown that it must be determined dynamically by all the contributing factors, the solid and fluid fractions, their velocities and  drifts. The same applies to the fluid erosion velocity $u_f^b$. This will have huge implications in the erosion-induced momentum productions which are dealt with at Section 4.2. As momentum productions play decisive role in the dynamics, mobility, destructive power, run-out and deposition morphology of mass flows (explained at Section 8), we must mechanically correctly describe the momentum productions with respect to the erosion velocities. This sheds light on the importance of the cross-mobility. This very crucial aspect, which was not realized before, and has been considered here for the first time. 
\\[3mm]
Figure \ref{Fig_5} shows the (strong) dominance of the cross-phase contributions over the phase-contributions on the solid and fluid erosion velocities. The cross-phase contributions are considered here for the first time revealing their essence in correctly describing the erosive mass transports. 

\subsection{Solid and fluid erosion rates}

 With the relations derived at Section 3.1, the total erosion rate $E$ in (\ref{Eqn_7}) is mechanically fully described.
 As the basal substrate, composed of the solid and fluid volume fractions $\alpha_s^b$ and $\alpha_f^b$, is entrained by the flow at the rate $E$, 
 the amounts produced by $\alpha_s^b E$ and $\alpha_f^b E$ are respectively added to 
 the solid and fluid components of the sliding mixture. 
 Since $\alpha_s^b  + \alpha_f^b  =1$, this facilitates constructing the solid erosion rate $\left ( E_s\right)$ and fluid erosion rate $\left ( E_f\right)$ by mechanically splitting the total erosion rate into the solid and fluid erosion rates  $\left (E = E_s + E_f\right)$ as:
\begin{equation}
E_s = \alpha_s^b E,\,\,\, E_f = \alpha_f^b E.
\label{Eqn_11}
\end{equation}
For this reason, we call $E$ in (\ref{Eqn_7}) the unified erosion rate for mixture mass flows.
This completes the derivation of the unified mechanical
erosion rate models for the solid and fluid components in two-phase mass flows.
Physical parameters and dynamical variables involved in erosion rates are explained at Section 6.4.
 
\section{Momentum productions for solid and fluid phases}

When the erosion induced produced (rate of) solid and fluid masses $E_s$ and $E_f$ are combined with the erosion velocities, the (rate of) momentum productions are obtained. There are two possibilities for this. Either, we consider the total erosion velocity $u^b$ and construct both the produced solid and fluid momenta $u^b E_s$ and $u^bE_f$. Or, the produced solid and fluid momenta $u_s^b E_s$ and $u_f^b E_f$ are constructed by utilizing the solid and fluid erosion velocities $u_s^b$ and $u_f^b$, respectively. 
The first choice is appropriate if at the time of erosion both the solid particle and fluid molecule from the bed move with similar velocities. This is relatively simple. However, if the eroded solid particle and the fluid molecule move with substantially different velocities, then the second choice is preferable as it can better describe the momentum productions.
Below, we consider the both of them as two alternative mechanical methods for the erosion-induced (rate of) momentum productions.

\subsection{In terms of the total erosion velocity}

First, we construct the momentum productions in terms of the down-slope ($x$-directional) component $u^b$ of the total velocity ${\bf u}^b$ of the eroded material, where, ${\bf u}^b = \left (u^b, v^b \right)$. With the solid and fluid erosion rates presented in (\ref{Eqn_11}), the solid and fluid momentum productions in the down-slope direction
$\mathcal M_{x_s}$ and $\mathcal M_{x_f}$ 
that enter the solid and the fluid momentum equations (Section 6, equations (\ref{Model_Final})) are $u^b E_s$ and $u^b E_s$, respectively: 
\begin{equation}
\mathcal M_{x_s} = u^b E_s = \lambda^b u^m E_s = \lambda^b u^m\alpha_s^b E = \lambda^b\alpha_s^b u^m E,
\label{Eqn_14u}
\end{equation}
\begin{equation}
\mathcal M_{x_f} = u^b E_f = \lambda^b u^m E_f= \lambda^b u^m\alpha_f^b E = \lambda^b \alpha_f^b u^m E,
\label{Eqn_15u}
\end{equation}
where the erosion drift relation $u^b = \lambda^b u^m$ has been employed from (\ref{Eqn_5}).
These momentum productions explicitly depend on the four aspects of flow: ($i$) erosion drift $\left (\lambda^b \right)$ characterizing the erosion velocity, ($ii$) volume fractions of solid and fluid in the erodible bed $\left (\alpha^b_{s}, \alpha^b_{f}\right)$, ($iii$) the flow velocity $\left(u^m\right)$, and ($iv$) the total erosion rate of the system $(E)$.
\\[3mm]
Similarly, the solid and fluid momentum productions $\mathcal M_{y_s}$ and $\mathcal M_{y_f}$  in the cross-slope ($y$) direction can be written, respectively, by consistently replacing $u$ by $v$ and $x$ by $y$ in (\ref{Eqn_14u})-(\ref{Eqn_15u}):
\begin{equation}
\mathcal M_{y_s} = v^b E_s = \lambda^b v^m E_s = \lambda^b v^m\alpha_s^b E = \lambda^b\alpha_s^b v^m E,
\label{Eqn_14v}
\end{equation}
\begin{equation}
\mathcal M_{y_f} = v^b E_f = \lambda^b v^m E_f= \lambda^b v^m\alpha_f^b E = \lambda^b \alpha_f^b v^m E.
\label{Eqn_15v}
\end{equation}
As in (\ref{Eqn_14u})-(\ref{Eqn_15u}), these momentum productions $\mathcal M_{y_s}$ and $\mathcal M_{y_f}$ analogously depend on the four aspects characterizing the flow mechanical properties.

\subsection{In terms of the solid and fluid erosion velocities}

\begin{figure}
\begin{center}
\includegraphics[width=9cm]{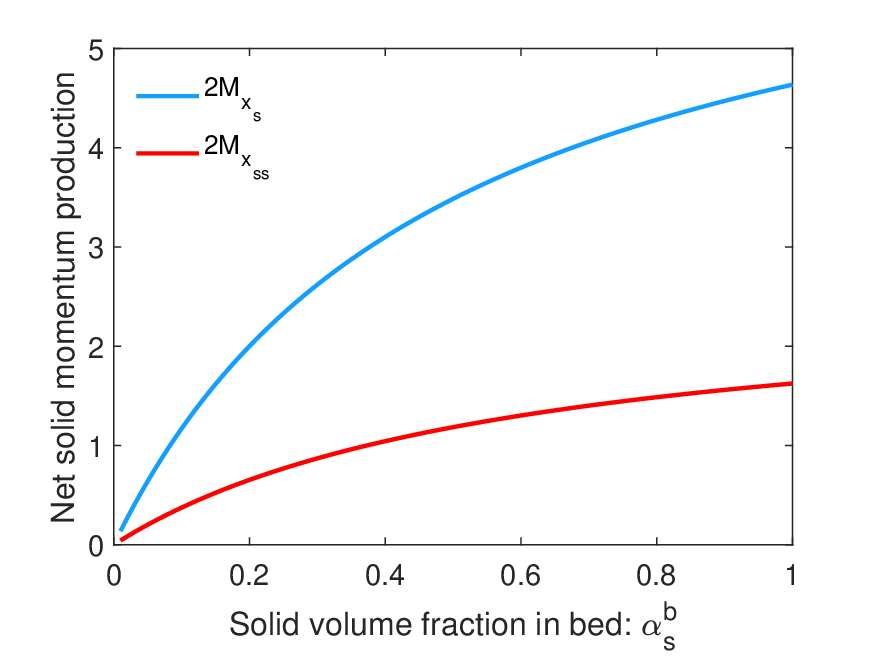}
\includegraphics[width=9cm]{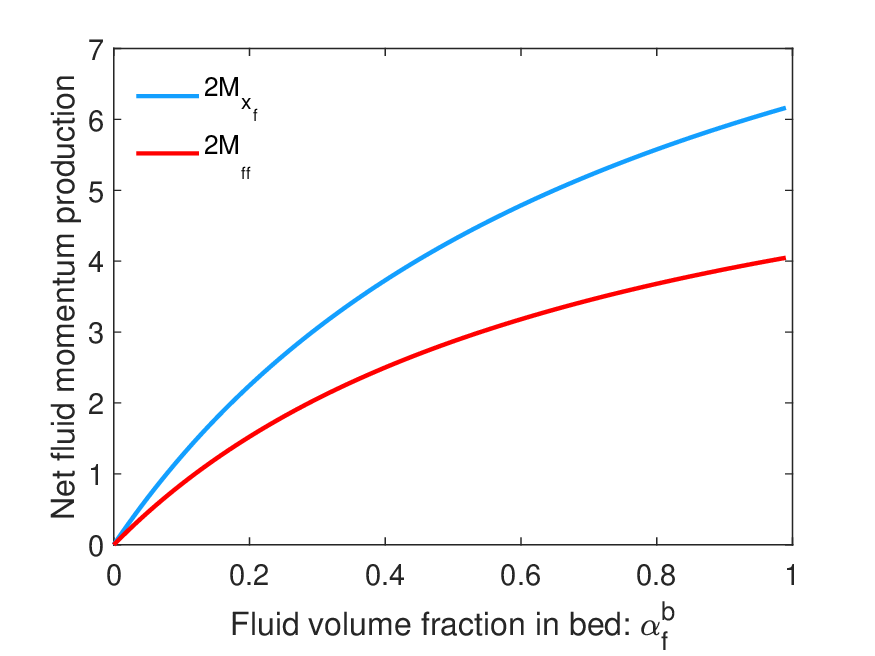}
  \end{center}
  \caption[]{Left: the net momentum production for solid, including both the phase- and cross-phase erosion velocity $\left (2 \mathcal M_{x_s}\right)$, and neglecting the cross-phase erosion velocity but only including the solid-solid phase erosion velocity $\left (2\mathcal M_{x_{ss}}\right)$ as given by (\ref{Eqn_12u}). 
  Right: the net momentum production for fluid, including both the phase- and cross-phase erosion velocity $\left(2 \mathcal M_{x_f}\right)$, and neglecting the cross-phase erosion velocity but only including the fluid-fluid- phase erosion velocity $\left(2\mathcal M_{x_{ff}}\right)$ as given by (\ref{Eqn_13u}).}
  \label{Fig_6}
\end{figure}

Next, we construct the $x$-directional solid and fluid momentum productions in terms of the solid and fluid erosion velocities.
With the velocities of the eroded solid particles and fluid molecules $u_s^b$ and $u_f^b$ from (\ref{Eqn_1NNN}), the solid and fluid momentum productions $\mathcal M_{x_s}$ and $\mathcal M_{x_f}$ that enter the $x$-directional solid and fluid momentum equations (Section 6) are $u_s^b E_s$ and $u_f^b E_s$, respectively. 
The momentum productions then take the forms:
\begin{equation}
\mathcal M_{x_s} = u_s^b E_s 
= \left[\alpha_s^m \lambda_{ss}^b u_s^m + \alpha_f^m \lambda_{fs}^b u_f^m\right] E_s
= \left[\alpha_s^m \lambda_{ss}^b u_s^m + \alpha_f^m \lambda_{fs}^b u_f^m\right] \alpha_s^b\, E,
\label{Eqn_12u}
\end{equation}
\begin{equation}
\mathcal M_{x_f} = u_f^b E_f 
= \left[ \alpha_s^m \lambda_{sf}^b u_s^m + \alpha_f^m \lambda_{ff}^b u_f^m\right] E_f 
= \left[ \alpha_s^m \lambda_{sf}^b u_s^m + \alpha_f^m \lambda_{ff}^b u_f^m\right] \alpha_f^b\, E. 
\label{Eqn_13u}
\end{equation}
It is important to note that these momentum productions explicitly depend on the five aspects of flow. These are:
($i$) erosion drifts $\left (\lambda^b_{ss}, \lambda^b_{fs}; \lambda^b_{sf}, \lambda^b_{ff}\right)$ characterizing erosion velocities, 
($ii$) volume fractions of solid and fluid in the landslide $\left (\alpha^m_{s}, \alpha^m_{f}\right)$, 
($iii$) volume fractions of solid and fluid in the erodible bed $\left (\alpha^b_{s}, \alpha^b_{f}\right)$, 
($iv$) $x$-directional solid and fluid velocities in the flow $\left(u^m_{s}, u^m_{f}\right)$, and 
($v$) the total erosion rate of the system $(E)$. 
However, $\lambda^b_{sf}$ and $\lambda^b_{ff}$ are already in $E$.
 Moreover, appearance of $\left (\alpha^m_{s}, \alpha^m_{f}\right)$ in (\ref{Eqn_12u})-(\ref{Eqn_13u}) is explicit, which was not the case in (\ref{Eqn_14u})-(\ref{Eqn_15u}). So, these momentum productions are broader than those in (\ref{Eqn_14u})-(\ref{Eqn_15u}). 
 \\[3mm]
 By neglecting the solid-fluid and fluid-solid interactions $\left (\lambda_{sf}^b = 0, \lambda_{fs}^b = 0\right)$, and considering only the solid-solid and fluid-fluid contacts between the solid and fluid across the erosion interface (so the factors $\alpha^m_s, \alpha^m_f$ do not appear, and $\lambda_{ss}^b, \lambda_{ff}^b$ become $\lambda_{s}^b, \lambda_{f}^b$), (\ref{Eqn_12u})-(\ref{Eqn_13u}) reduce to the solid and fluid momentum productions in Pudasaini and Fischer (2020a), which, however, are largely incomplete. 
 \\[3mm]
The major role the erosion velocities play are in constituting momentum productions, which in turn, rule the entire dynamics of erosive mass flows. 
Field measurements have shown that the erosion rate $E$ can vary in the range 0.002-0.8 m$s^{-1}$, but can also exceed 1.0  m$s^{-1}$ (Berger et al., 2011; Iverson et al., 2011; McCoy et al., 2012). So, for the demonstrative purpose, we take a plausible value as $E = 0.25$ m$s^{-1}$. 
Figure \ref{Fig_6} displays the complete net momentum production for solid including both the contributions in the solid erosion velocity as mobilized by the solid and fluid from the landslide, $2 \mathcal M_{x_s}$, and the incomplete net momentum production for solid that only includes the contribution in the solid erosion velocity as mobilized by the solid from the landslide, $2\mathcal M_{x_{ss}}$. Similar analysis holds for fluid net momentum productions. These results reveal the need of including both the phase- and cross-phase momentum productions as the differences are substantial. This implies that the erosion-induced net momentum productions can be closer to the gravity loads on the landslide. So, the erosive mass transports must include the complete descriptions of erosion velocities and their full involvement in net momentum productions.
 \\[3mm]
 In general it is less likely, but, in special situation the difference between the solid and fluid phase velocities in the landslide may be negligible, i.e., $u_s^m \approx u_f^m =: u^m$. Even then, the solid and fluid in the landslide may differently mobilize the solid in the bed, and so, $\lambda_{ss}^b \neq \lambda_{fs}^b$. But, if the solid and fluid in the landslide may similarly mobilize the solid in the bed, only then, $ \lambda_{ss}^b \approx \lambda_{fs}^b =: \lambda^b$. In this situation, (\ref{Eqn_12u}) is $\lambda^b\alpha_s^b u^m E$, which is (\ref{Eqn_14u}). Similar analysis applies to (\ref{Eqn_13u}). This implies the wide spectrum and the physically fully consistent modelling of momentum productions in (\ref{Eqn_12u})-(\ref{Eqn_13u}) associated with the solid and fluid erosion velocities.
\\[3mm]
Similarly, the $y$-directional momentum productions (in terms of the corresponding solid and fluid erosion velocities: $v^b_s, v^b_f$) are obtained by consistently replacing $u$ by $v$ and $x$ by $y$ in (\ref{Eqn_12u})-(\ref{Eqn_13u}):
\begin{equation}
\mathcal M_{y_s} = v_s^b E_s 
= \left[\alpha_s^m \lambda_{ss}^b v_s^m + \alpha_f^m \lambda_{fs}^b v_f^m\right] E_s
= \left[\alpha_s^m \lambda_{ss}^b v_s^m + \alpha_f^m \lambda_{fs}^b v_f^m\right] \alpha_s^b\, E,
\label{Eqn_12v}
\end{equation}
\begin{equation}
\mathcal M_{y_f} = v_f^b E_f 
= \left[ \alpha_s^m \lambda_{sf}^b v_s^m + \alpha_f^m \lambda_{ff}^b v_f^m\right] E_f 
= \left[ \alpha_s^m \lambda_{sf}^b v_s^m + \alpha_f^m \lambda_{ff}^b v_f^m\right] \alpha_f^b\, E. 
\label{Eqn_13v}
\end{equation}
As in (\ref{Eqn_12u})-(\ref{Eqn_13u}), these momentum productions also explicitly depend on the five aspects of flow: 
($i$) erosion drifts $\left (\lambda^b_{ss}, \lambda^b_{fs}; \lambda^b_{sf}, \lambda^b_{ff}\right)$ characterizing erosion velocities, 
($ii$) volume fractions of solid and fluid in the landslide $\left (\alpha^m_{s}, \alpha^m_{f}\right)$,
($iii$) volume fractions of solid and fluid in the erodible bed $\left (\alpha^b_{s}, \alpha^b_{f}\right)$, 
($iv$) $y$-directional solid and fluid velocities in the flow $\left(v^m_{s}, v^m_{f}\right)$, and 
($v$) the total erosion rate of the system $(E)$.
However, $\lambda^b_{sf}$ and $\lambda^b_{ff}$ are already in $E$.
In total, there are five erosion drifts in (\ref{Eqn_12u})-(\ref{Eqn_13v}) including those in $ E$, that are needed in application, namely, $\lambda^b; \lambda_{ss}^b, \lambda_{fs}^b, \lambda_{sf}^b, \lambda_{ff}^b$, as explained at Section 3.1.3.,  all of which are explicitly known with the bed inertial numbers $N^I, N_{ss}^I, N_{fs}^I, N_{sf}^I, N_{ff}^I$.
\\[3mm]
By comparing (\ref{Eqn_14u})-(\ref{Eqn_15u}) with (\ref{Eqn_12u})-(\ref{Eqn_13u}), it is interesting to observe that the complex phase erosion velocities
$\left[\alpha_s^m \lambda_{ss}^b u_s^m + \alpha_f^m \lambda_{fs}^b u_f^m\right]$ and
$\left[ \alpha_s^m \lambda_{sf}^b u_s^m + \alpha_f^m \lambda_{ff}^b u_f^m\right]$
are intrinsically related to the structurally simple total erosion velocity $\lambda^b u^m$. Similar situation applies between (\ref{Eqn_14v})-(\ref{Eqn_15v}) and 
(\ref{Eqn_12v})-(\ref{Eqn_13v}). The momentum productions (\ref{Eqn_14u})-(\ref{Eqn_15v}) are structurally easier  over (\ref{Eqn_12u})-(\ref{Eqn_13v}). 
However, whether (\ref{Eqn_14u})-(\ref{Eqn_15v}) or (\ref{Eqn_12u})-(\ref{Eqn_13v}) are more appropriate should be verified with applications to erosive laboratory flows and real events. 

\subsection{The erosion matrix}

The structures of erosion velocities in (\ref{Eqn_12u})-(\ref{Eqn_13u}) can be written in a compact matrix-vector form:
\begin{equation}
{\boldsymbol{S}}_{_E}^b  {\boldsymbol{A}}^m {\bf u}^m = {\bf u}^b,
\label{Eqn_SE}
\end{equation}
where 
\begin{equation}
{\boldsymbol{S}}_{_E}^b  = 
\begin{bmatrix}
\lambda_{ss}^b & \lambda_{fs}^b \\[2mm]
\lambda_{sf}^b & \lambda_{ff}^b 
\end{bmatrix}
, \,\, 
\boldsymbol{A}^m  = 
\begin{bmatrix}
\alpha_s^m & 0 \\[2mm]
0 & \alpha_f^m 
\end{bmatrix}
, \,\, 
{\bf u}^m = 
\begin{bmatrix}
u_s^m \\[2mm]
u_f^m
\end{bmatrix}
, \,\, 
{\bf u}^b = 
\begin{bmatrix}
u_s^b \\[2mm]
u_f^b
\end{bmatrix},
\label{Eqn_SE1}
\end{equation}
 are the matrix of the (phase-phase and cross-phase) erosion drifts, diagonal matrix of the volume fractions in the landslide, the vector of flow velocities in the landslide, and the vector of erosion velocities at the bed. we call ${\boldsymbol{S}}_{_E}^b$ the (system) erosion matrix. This is invented here. 
 Note that ${\boldsymbol{S}}_{_E}^b$ and $\boldsymbol{ A}^m$ can be combined to obtain another matrix, say, ${\boldsymbol{ S}}_{_{E_e}}^b = {\boldsymbol{ S}}_{_E}^b \boldsymbol{ A}^m$. We call it the effective erosion matrix.
 There are some interesting and important properties of the erosion matrix. For the vanishing off-diagonal elements ${\boldsymbol{ S}}_{_E}^b$ degenerates to imply the simple erosion velocities in Pudasaini and Fischer (2020a), but without the cross-phase interactions, which is not complete. The determinant of ${\boldsymbol{ S}}_{_E}^b$ shows that ${\boldsymbol{ S}}_{_E}^b$ is primarily ruled by the solid and fluid masses in the flow, $\left (\rho_s^m\alpha_s^m\right)$ and $\left (\rho_f^m\alpha_f^m\right)$, appearing in its numerator. Which means, erosion is essentially driven by the flow dynamics, which can be easily understood. However, the intensity of erosion depends on the  strength of the resisting bed material as ${\boldsymbol{ S}}_{_E}^b$ contains the factor 
 $1/\left (\rho_s^m\alpha_s^m + \rho_s^b\alpha_s^b\right)
    \left (\rho_f^m\alpha_f^m + \rho_f^b\alpha_f^b\right) 
 - 1/\left (\rho_s^m\alpha_s^m + \rho_f^b\alpha_f^b\right)
    \left (\rho_f^m\alpha_f^m + \rho_s^b\alpha_s^b\right)$. 
    Since for mixture flows $\left (\rho_s^m\alpha_s^m\right)$ and $\left (\rho_f^m\alpha_f^m\right)$ are finite positive values, and 
    $\left (\rho_s^m\alpha_s^m - \rho_f^m\alpha_f^m\right)$ and 
    $\left (\rho_s^b\alpha_s^b - \rho_f^b\alpha_f^b\right)$ substantially deviate away from zero,
    ${\boldsymbol{ S}}_{_E}^b$ is non-singular and well defined, so is $\boldsymbol{ A}^m$. 
    This means the system (\ref{Eqn_SE}) is invertible, allowing to reconstruct the flow velocities of the landslide from the knowledge of the erosion velocities, a novel perception in erosive mass transports.
    Moreover, this also tells that the erosion intensity increases as the solid and fluid phase masses in the flow and the bed deviate from one another. 
    Hence,  ${\boldsymbol{ S}}_{_E}^b$ characterizes the erosion mechanism of the system. 
    The erosion matrix formally maps the flow dynamics (volume fractions and velocities of the flow) to the erosion velocities of the bed.
    This shows that, as the erosion velocity ${\bf u}^b$ governs the system, the process of erosion is jointly determined by the erosion matrix, volume fractions of the flow and the flow velocities. So, we have presented the first systematic, compact and the complete description of the dynamics and mechanical process of erosion.
 
 \section{Comprehensive modelling of erosion including particle collisional stress}

 Often as the dispersed regions (e.g., fronts) in granular and debris flows do not sustain the lasting contacts among the grains, the flow induced frictional shear stress on the erodible bed can be insignificant. In this situation, the collisional stress plays a significant to commanding role in mobilizing the bed material in erosional flows. Evidently, observations show that significant erosion can still take place even when the shear strength of the bed material is larger than the frictional shear stress induced by the flow as erosion may predominantly take place, for example, as the flow front, where most collisional stresses are generated, passes over some local regions (McArdell et al., 2007; Berger et al., 2011; de Haas and Woerkom, 2016; Song and Choi, 2021). This suggests the need to complement the frictional shear stress with the collisional shear stress in constituting the comprehensive modelling of erosive mass flows.  

\subsection{Complementing the frictional stress with the collisional stress in particle motion}

In general, the granular and debris flows may be best described by a combination of the essentially two distinct rheological behaviors, the frictional and the collisional. The previous descriptions in Section 2 – Section 4 are developed solely for the frictional granular flows. Below, we also present modelling frame for collisional stress generations and combine it with the frictional stresses and implement these in the unified modelling resulting in the comprehensive description of erosional granular (debris) mass transports. 
\\[3mm]
The collisional stress (or the fluidization) in the granular (debris) flows are often described by the Bagnold grain-inertial stress (Bagnold, 1954; Campbell, 2006; Hsu et al., 2008; Pudasaini, 2011; Yohannes et al., 2012):
\begin{eqnarray}
\begin{array}{lll}
\displaystyle{
\tau_{B,ss}^m = \rho_s^m\lambda_m^2 d_m^2\left(\frac{\partial u_s^m}{\partial z}\right)^2 \alpha_s^m,}
\label{Eqn_Col1}
\end{array}    
\end{eqnarray}
where, $d_m$ is the representative particle diameter, and $\lambda_m$, as Bagnold explained, is the linear concentration of the solid particles that (by assuming $\alpha_s$ as the volumetric concentration) can be represented by
\begin{eqnarray}
\begin{array}{lll}
\displaystyle{
\lambda_m = \left[ \left( \frac{\alpha^m_{s,max}}{\alpha_s^m}\right)^{1/3}-1\right]^{-1},}
\label{Eqn_Col2}
\end{array}    
\end{eqnarray} 
with the maximum possible particle concentration ${\alpha_{s, max}^m}$. 
$\lambda_m$ plays a key role in the description of the collisional regime as $\lambda_m$ is very sensitive to $\alpha_s^m$. There are some loose suggestions (Iverson, 1997) to replace $\lambda_m^2$ with $\alpha_s^m$. However, in contrast to Bagnold, it appears that those suggestions are made without identifying the proportionality factor in explaining the grain-inertial stress. So, analyses based on such simplifications may raise concern.  
\\[3mm]
The shear rate ${\partial u_s^m}/{\partial z}$ in (\ref{Eqn_Col1}) can be estimated by the ratio between the flow velocity $u_s^m$ and the flow depth $h$ (Savage and Hutter, 1989). Then, (\ref{Eqn_Col1}) reduces to:
\begin{eqnarray}
\begin{array}{lll}
\displaystyle{
\tau_{B,ss}^m = \rho_s^m\lambda_m^2 d_m^2\left(\frac{u_s^m}{h}\right)^2\alpha_s^m,}
\label{Eqn_Col3}
\end{array}    
\end{eqnarray} 
where, $u_s^m$ now is the velocity difference across the flow (or, shear-layer) and $h$ is the flow (or, shear-layer) thickness.
 This shows that, collisional stress is much higher for flows with: (i) larger particles, (ii) thin flows, and (iii) higher velocities. This is in contrast to the frictional (Coulomb) stress, which is higher for think, frictional flows.
\\[3mm]
So, the total solid stress (sum of the frictional $\tau^m_{C,ss}$ as modelled in (\ref{Eqn_1NN}), and the collisional $\tau^m_{B,ss}$ in (\ref{Eqn_Col3})) for the particles in the flow becomes
\begin{eqnarray}
\begin{array}{lll}
\displaystyle{\tau^m_{ss} = \left(1-\gamma^m\right) \rho^m_s g^z h\mu^m _s\alpha^m _s 
               +  \rho^m_s\lambda_m^2 d_m^2\left(\frac{u_s^m}{h}\right)^2\alpha_s^m,
}
\label{Eqn_Col6}
\end{array}    
\end{eqnarray} 
where, unlike in many previous considerations (Iverson, 1997; de Haas and Woerkom, 2016; Song and Choi, 2021; Li et al., 2024; Li et al., 2025) the slope normal component of gravity acceleration, $g^z$, appears consistently.

\subsection{Dominant rheology}

  As explained, there are two types of stress generating mechanisms in granular (debris) flows: The Coulomb frictional stress that depends on the normal load combined with the grain frictional coefficient (independent of the shear-rate), and the Bagnold grain-inertial stress that quadratically depends on the shear-rate (but, independent of the normal load). Which one of these fundamentally different rheological behaviors prevails and dominates the granular flows is a great question, but is required in properly describing such complex flows. However, there is no clear distinction between these two rheological states, or the flow regimes. Yet, based on the Bagnold’s grain-inertial stress, Savage and Hutter (1989) put forward the concept of taking the ratio between the collisional stress and the total normal stress generated by the granular flows to characterize the importance of collisional stresses over the normal stress. Following this, and our stress definition from (\ref{Eqn_Col6}), the ratio reads:
\begin{eqnarray}
\begin{array}{lll}
\displaystyle{R_{SH} 
= \frac{\rho^m_s\lambda_m^2 d_m^2\D{\left(\frac{u_s^m}{h}\right)^2}\alpha^m_s} {\left(1-\gamma_s^m\right)\rho^m_s g^z h\mu^m_s\alpha^m_s + \rho^m_s\lambda_m^2 d_m^2\D{\left(\frac{u_s^m}{h}\right)^2}\alpha^m_s}.                               
}
\label{Eqn_Col7}
\end{array}    
\end{eqnarray} 
Savage and Hutter (1989) rudimentarily suggested that values of $R_{SH}$ greater than 0.1 may imply the dominant collisional stress over the frictional stress. Some (Stock and Dietrich, 2006; Hsu et al., 2008) also considered (\ref{Eqn_Col7}) as legitimate representation in distinguishing the frictional or collisional flow regimes. 
\\[3mm]
However, in the literature, almost exclusively (Iverson, 1997; Stock and Dietrich, 2006; de Haas and Woerkom, 2016; Baselt et al., 2021;  Li et al., 2024; Li et al., 2025), (\ref{Eqn_Col7}) is used differently as the ratio between the collisional stress and the Coulomb stress (but not the total stress as suggested by Savage and Hutter (1989)), 
 but, still use the Savage and Hutter (1989) criterion to distinguish the frictional or collisional dominance in the flow. So, the conclusion drawn based on this
cannot properly tell us which of the stresses, the frictional or the collisional, dominates the flow regimes, or both of them are significantly contributing  to the momentum transfer during the granular flows. 

\subsection{A different perspective}

Structurally, (\ref{Eqn_Col7}) says that very small (not just smaller that 0.1) values of $R_{SH}$ correspond to a dominance of stresses generated by Coulomb frictional interactions, and that larger values (close to unity, evidently not just greater than 0.1) of $R_{SH}$ basically correspond to the situation generating collisional stresses. This raises an anxiety: what is the more acceptable way to delineate the frictional and collisional regimes. One could imagine that the natural way to delineate these flow regimes should be based on the magnitude of the ratio of these stresses. 
 So, to better represent the flow regimes, we propose that
 the general hypothesis should be to determine the dominance of the flow regimes not by looking at the exaggerate ratio between the collisional stress to the total stress, but rather by simply directly looking at the ratio between the collisional and the frictional stresses as:
\begin{eqnarray}
\begin{array}{lll}
\displaystyle{R_P 
= \frac{\rho^m_s\lambda_m^2 d_m^2\D{\left(\frac{u_s^m}{h}\right)^2}\alpha_s^m} {\left(1-\gamma_s^m\right) \rho^m_s g^z h\mu^m_s\alpha^m_s}.                        
}
\label{Eqn_Col9}
\end{array}    
\end{eqnarray}
Then, it is logical and legitimate to perceive that if  $ R_P \ll 1$, the flow is friction-dominated, 
 and if $ R_P \gg 1$, the flow is collision-dominated.
However, (\ref{Eqn_Col9}) also cannot furnish any hard-rule in distinguishing these two regimes. This only provides some blurred picture. What we can say about is the description of any model by combining the two flow laws as:
\begin{eqnarray}
\begin{array}{lll}
\displaystyle{\tau^m_{ss}: = C_s^m\left(1-\gamma^m\right) \rho^m_s g^z h\mu^m _s\alpha^m _s 
               +  \left(1-  C_s^m\right)\rho^m_s\lambda_m^2 d_m^2\left(\frac{u_s^m}{h}\right)^2\alpha_s^m,
}
\label{Eqn_Col6_a}
\end{array}    
\end{eqnarray} 
and judge their mechanical dominance by looking on the model performance with the dynamics, run-out and deposition of the considered flows in the laboratory, or in the field. In (\ref{Eqn_Col6_a}), $C_s^m \in [0, 1]$ linearly combines the contributions due to the frictional and collisional stresses determining the total stress generated by the granular flow. As $C_s^m \to 1$ the flow becomes friction-dominated, and as $C_s^m \to 0$ the flow turns to be collision-dominated. However, in general, we suggest to properly use $C_s^m \in [0, 1]$. 

\subsection{Generating collisional flows}

 There is no common rule to designate the collisional flows. In general, lower concentrations and higher shear rates may be regarded as being in the grain inertia regime as collisional stresses can be dominant in this situation (Savage and Sayed, 1984; Hanes and Inman, 1985). For relatively thin-layer flows, sufficiently high bed slope to generate the grain inertia dominant regime as fluidization may prevail (Savage and Hutter, 1989), most probably in the flow front. However, it also depends on the particle size and density as in such dispersed situation the drag may play a major role. Yet, looking on the structure of the stress generating mechanisms, for relatively rapid, thin flows down rough steep slopes with moderate particle concentration we may assume collisional flows, and frictional flows in the opposite situation.

 \subsection{Stresses at the erodible bed}
 
As for the flow, we also need to describe the collisional stress for the particles in the bed mobilized by the erosive flows. However, it requires the proper understanding of the erosion-entrainment mechanism and the associated dynamics of the bed particles agitated by the collisional impacts from the flow. Following the Coulomb-type frictional stress and the Bagnold-type collisional stresses as designed above, these two basal stresses can be written as:
\begin{eqnarray}
\begin{array}{lll}
\displaystyle{\tau^b_{C,ss} =\left(1-\gamma_s^b\right)\rho^b_s g^z h\mu^b_s\alpha_s^b,\,\,\,\,\,
\tau^b_{B,ss}  = \rho^b_s\lambda_b^2 d_b^2\left(\frac{u_s^b}{h}\right)^2\alpha_s^b.                        
}
\label{Eqn_Col10}
\end{array}    
\end{eqnarray}
Then, the total solid stress at the mobilized erodible bed is given by:
\begin{eqnarray}
\begin{array}{lll}
\displaystyle{\tau^b_{ss} =  \left(1-\gamma_s^b\right)\rho^b_s g^z h\mu_s^b\alpha^b_s
                 +  \rho^b_s\lambda_b^2 d_b^2\left(\frac{u_s^b}{h}\right)^2\alpha_s^b.                    
}
\label{Eqn_Col11}
\end{array}    
\end{eqnarray}
  The way of selecting the Coulomb-type frictional stress or the Bagnold-type collisional stress at the bed follows as in the mixture. Moreover, as for the total solid stress in the flow, the total solid stress at the bed can also be considered as a linear combination between the frictional and the collisional stresses, with the definition $C_s^b \in [0, 1]$, as:
  \begin{eqnarray}
\begin{array}{lll}
\displaystyle{\tau^b_{ss}: =  C_s^b\left(1-\gamma_s^b\right)\rho^b_s g^z h\mu_s^b\alpha^b_s
                 +  \left(1- C_s^b\right)\rho^b_s\lambda_b^2 d_b^2\left(\frac{u_s^b}{h}\right)^2\alpha_s^b,                    
}
\label{Eqn_Col11_a}
\end{array}    
\end{eqnarray}
  where, as $C_s^b \to 1$ the bed stress is in the friction-dominated state, and as $C_s^b \to 0$ the bed turns to be in the collision-dominated state.

\subsection{Comprehensive model}  
  
With these considerations for the frictional and collisional stresses in the flow and in the bed, in situations when the collisional stress can be non-negligible, all the Coulomb frictional stresses at Section 2 – Section 4 at the flow and the bed must be replaced by the total solid stresses (\ref{Eqn_Col6}) and (\ref{Eqn_Col11}); equivalently (\ref{Eqn_Col6_a}) and (\ref{Eqn_Col11_a}); particularly at (\ref{Eqn_3N}) or at (\ref{Eqn_3al}). 
Moreover, the solid stresses in the solid source terms in the momentum balance equations must also be amended accordingly (see, Section 7). 
This will eventually complete the comprehensive model development for the erosional mass flows that accommodates both the frictional and collisional stress-generating mechanisms at the flow and at the erosional bed. Such a unified and comprehensive modelling of erosive multi-phase mass flow is novel.  
 
\section{Essence of unified mechanical erosion rates and momentum productions}

In the form the erosion rate models developed here are similar to those already 
presented in Pudasaini and Fischer (2020a) which offers the basic mechanical foundation. We have utilized three important
mechanical aspects from Pudasaini and Fischer (2020a) in constructing the new erosion
 rate models. These are: the jump in the shear stresses and the momentum fluxes 
 across the erosion-interface, the shear velocity and the erosion drifts.
 However, there are substantial differences between the erosion rate models in 
 Pudasaini and Fischer (2020a) and the ones developed here. 
 Moreover, we have presented extensive and complete multi-phase momentum productions than those in Pudasaini and Fischer (2020a).
 We discuss several
 crucial properties of the new modelling framework.
 
 \subsection{Novelty and significance of the new approaches for shear stresses and interactions}

The general situation of the stress jump given by (\ref{Eqn_3N}), and the eight different scenarios (\ref{Eqn_4N})-(\ref{Eqn_9N}), demonstrate its richness, the spectrum of applicability, and urgency of the new unified, consistent and comprehensive multi-phase erosion model presented here. The reductions at Section 2.5 also signify that the interacting stress structures in (\ref{Eqn_1N}) and (\ref{Eqn_2N}) are mechanically valid. Many of these aspects could not be described by any of the existing mass flow models. The professional and engineers will find the model structure (\ref{Eqn_3N}) and their reductions (\ref{Eqn_4N})-(\ref{Eqn_9N}) intuitive and useful in solving applied mass flow simulation associated with erosive events.
\\[3mm]
In the Pudasaini and Fischer (2020a) two-phase mechanical erosion model, the fluid-solid ($fs$) and solid-fluid ($sf$) interaction shear stresses (\ref{Eqn_2NN}) do not exist as the cross-shear stresses, but rather were directly used as $\tau_{ff}^m$ and $\tau_{ff}^b$ with the classical Chezy-frictions, which are physically inconsistent for their true cross-phase interactions. Moreover, in Pudasaini and Fischer (2020a), the fluid-fluid ($ff$) shear stresses (\ref{Eqn_3NN}) do not exist at all. We have made a fundamental advancement in modelling multi-phase erosive mass transport with entirely novel and innovative mechanical ideas for the shear stresses associated with multi-phase erosive mass transports with all possible interactions and natural shear stress descriptions for these interactions, 
 including the comprehensive description of the solid particles with the unified representation of the frictional and collisional stresses.
There are several important aspects of the new unified modelling of the multi-phase erosive mass transport. 
\\[3mm]
{\bf I.} As shown in (\ref{Eqn_1N}) and (\ref{Eqn_2N}), there are four fundamentally different types of interactions between the solid and fluid in the landslide and the solid and fluid in the bed material. These are: 
the solid-solid interaction 
$\alpha_f^b\left [ \tau_{ss}^m - \tau_{ss}^b \right ]$, 
which means solid in the landslide applies shear stress to the solid in the bed that is resisted by the solid material from the bed. Similar situations apply to other interactions: 
the solid-fluid 
$\alpha_s^b\left [ \tau_{sf}^m - \tau_{sf}^b \right ]$, 
fluid-solid:
$\alpha_s^b\left [ \tau_{fs}^m - \tau_{fs}^b \right ]$, and 
fluid-fluid:
$\alpha_f^b\left [ \tau_{ff}^m - \tau_{ff}^b \right ]$ interactions. 
The only two-phase erosion model that exists is by Pudasaini and Fischer (2020a). However, they only consider  solid-solid 
 $\left [ \tau_{ss}^m - \tau_{ss}^b \right ] = \left [ \tau_{s}^m - \tau_{s}^b \right ]$
and fluid-fluid 
$\left [ \tau_{ff}^m - \tau_{ff}^b \right ] = \left [ \tau_{f}^m - \tau_{f}^b \right ]$
 interactions, respectively, for which only the single suffix $s$ and $f$ are used as there are no cross-phase interactions. These are just the direct solid-solid, and fluid-fluid interactions. Moreover, therein, fluid-fluid interactions are modelled with the classical Chezy-type frictions, which, as stated at Section 2.4.3, are mechanically inconsistent. 
\\[3mm]
{\bf II.} Here, for the first time, we introduced mechanically important inter-phase solid-fluid 
 $\alpha_s^b\left [ \tau_{sf}^m - \tau_{sf}^b \right ]$ (second element on the left hand side of (\ref{Eqn_1N}))
and fluid-solid 
$\alpha_s^b\left [ \tau_{fs}^m - \tau_{fs}^b \right ]$ (first element on the left hand side of (\ref{Eqn_2N}))
interactions across the erosion-interface.
\\[3mm]
{\bf III.} The shear applied by the fluid from the flow to the solid at the bed
$\tau_{fs}^m$ in (\ref{Eqn_2NN})
 follows the usual Chezy-type friction with the fluid velocity at the base of the landslide. However, surprisingly, we revealed that the shear resistance by the fluid at the bed against the applied shear from the solid from the flow
$\tau_{sf}^b$ in (\ref{Eqn_2NN})
 does not directly follow the usual Chezy-type friction with the fluid velocity at the bed, rather it is associated with the solid velocity at the base of the landslide. This is so, because this type of basal shear stress is induced by the shear load of the solid particles from the landslide applied to the fluid at the bed. This also led to a challenge in constructing the cross-erosion drift $\lambda_{sf}^b$, which, as discussed at Section 2.4.2, required a different type of erosion drift relationship between the fluid in the bed and the solid in the landslide. As revealed at Section 3.1.3, this is a novel understanding.
\\[3mm]
{\bf IV.} The further astonishing fact we discovered here is the fluid-fluid 
 $\alpha_f^b\left [ \tau_{ff}^m - \tau_{ff}^b \right ]$
interaction. One may simply think to directly apply the classical Chezy-type relation as in Pudasaini and Fischer (2020a) (which, as discussed above, is not fully consistent), or simple use of viscous fluid shear stresses for this interaction. However, the situation of multi-phase erosive mass transport is different, and requires physically meaningful complex mechanical interfacial shear stresses. The point is that, these are fluid-fluid interactions, so Chezy-type friction is not applicable. Moreover, the fluids both in the landslide and basal materials are not the free fluids that can directly interact to each other and follow the simple rule of viscous shear stress. These fluids are contained inside the matrices of the solid particles in the landslide and the solid particles in the bed material. So, in addition to its own viscosity and velocity, the shear stress of the fluid in the landslide is influenced by the geometrical and mechanical properties of the basal material, particularly the permeability of the bed. The same applies to the shear resistance of the fluid in the erodible bed. As shown in (\ref{Eqn_3NN}), this required novel descriptions of the fluid-fluid interactions at the interface. 
\\[3mm]
These four aspects clearly manifest the physical novelty and significance of the modelling approach 
for the shear stresses and interactions between different materials across the landslide-bed interface,
presented here for the complex multi-phase erosive mass transports that are still lacking.
 
 \subsection{Advantages of the unified mechanical erosion rate model}
 
 One of the main purposes of this contribution is to construct a unified and physically consistent mechanical erosion rates for multi-phase mass flows.
  It takes into account all the frictional, collisional and viscous stress generating mechanisms in a comprehensive way.
 For several reasons, the novel unified multi-phase mechanical erosion models (\ref{Eqn_7}) or (\ref{Eqn_11}) are required.
 The total basal erosion rate $E$ in (\ref{Eqn_7}) is the consistent and exact sum of the solid and fluid erosion rates $E_s$ and $E_f$ in (\ref{Eqn_11}), i.e., $E = \alpha_s^b E + \alpha_f^b E$. 
 This is crucial. We have presented the first such mechanical model for mixture mass flows. 
 These erosion rates inherently contain the solid and fluid fractions of the bed material $\alpha_s^b$ and $\alpha_f^b$, respectively.
 These solid and fluid erosion rates $E_s = \alpha_s^b E$ and $E_f = \alpha_f^b E$ consistently take the solid and fluid fractions $\alpha_s^b$ and $\alpha_f^b$ from the bed and customarily supply them to the flow. Now, from the total eroded material, $\alpha_s^b$ and $\alpha_f^b$ fractions are persistently incorporated into the solid and fluid components in the moving material. 
 This was not possible with the existing multi-phase erosion models (Pudasaini and Fischer, 2020a). 
 That could only be achieved partially, but not consistently, by manually adjusting several parameters including the solid and fluid erosion drifts, Chezy friction coefficients, and the shear velocity factor.
 So, in the existing models, the sum of the solid and fluid erosion rates do not realistically correspond to the eroded material from the bed. 
 Moreover, for fast flows, the fluid erosion rate can be unrealistically higher than the solid erosion rate which poses a great problem in consistent  selection of the parameters in the erosion rates in Pudasaini and Fischer (2020a) model. 
 Thus, natural erosion rates that correspond to the events cannot be obtained from the solid and fluid erosion rates developed by Pudasaini and Fischer (2020a). Here, we have addressed this great existing problem in erosive mass transport as required by practitioners 
 with a unified erosion rate model such that the sum of the derived solid and fluid erosion rates automatically satisfies the natural criterion without any adjustment. 
 \\[3mm]
 The previous explicit erosion rate models can be applied to the situations when both the flowing mixture and base material include sufficient amount of solid and fluid phases. However, in certain situations, e.g., if the flowing material is almost a fluid, then solid material from the bed cannot be entrained into the flow. The newly developed unified erosion rate model overcomes this limitation. 
 \\[3mm]
 The new unified erosion rate models include all interactions between the solid and fluid phases in the landslide and the bed across the erosion interface. Each of these interactions are mechanically and dynamically important. 
 However, not all of them were recognized previously, fiercely limiting the applicability of the existing erosion models. 
 We considered all phase-phase and cross-phase interactions across the flow-bed interface. 
 We removed several shortcomings inherent in the existing erosion rate modelling (Pudasaini and Fischer, 2020a), where 
 the fluid-fluid interactions were not described mechanically appropriately.
 Moreover, the true solid-fluid and fluid-solid interactions were ignored previously.
 These complex interactions were beyond our present understanding. We solved these problems by presenting novel, mechanical shear stress models for the solid-fluid and fluid-fluid interactions.
 \\[3mm]
 Surprisingly, erosion velocities for both the solid and fluid appeared to be extensive.  
 For the first time, we revealed the complex, but real mechanical situations that the solid particle at the bed is pushed and then mobilized both by the solid and fluid in the flow. However, the existing model (Pudasaini and Fischer, 2020a) only considered the mobilization of the basal solid particle by the solid from the landslide, but ignored its mobilization by the fluid from the landslide. This also applies to the erosion velocity of the fluid molecule at the bed. These realizations are crucial with which we presented mechanically complex, extended erosion velocity (Section 2.3) and mobility (Section 8) models by incorporating all the interactions between the particles and fluids across the erosion-interface. 
 We constructed a novel, unified and complete mechanical mass and momentum production rates (Section 3.4 and Section 4) and embedded them into the dynamical model equations (mass and momentum balances, Section 7). This overcomes the severe limitations of existing erosion model, and 
 opens a wide spectrum of possibilities for real applications in complex mass flow simulations. 

\subsection{Wide applicability of the new model}
 
One of the most important aspects of the new erosion rate model presented here
is that it can be applied to any flow situations and the bed morphologies,
irrespective of the number of contributory components in the flow and in
the bed. 
As exclusively discussed at Section 2.4 and Section 2.5, the new erosion rate models can be applied to nine different scenarios: 
($i$) sliding debris mixture entraining different debris mixture from the bed, 
($ii$) dry landslide entraining dry bed,
($iii$) fluid flow entraining fluid material from bed - river, lake or reservoir fluid,
($iv$) dry landslide entraining river, lake or reservoir fluid,
($v$) flood entraining soil, sand or gravel from the bed,
($vi$) debris flow entraining soil, sand or gravel from the bed,
($vii$) debris flow entraining fluid material from bed - river, lake or reservoir fluid,
($viii$) dry landslide entraining soil, sand or gravel from the bed, and
($ix$) flood entraining debris mixture from the bed.
Such a wide
spectrum of erosion modelling is presented here for the first time with
the unified modelling approach. The new model is fully mechanical. So, the model presents a great opportunity in
solving different applied, technical, engineering and geomorphological
problems.

\subsection{Parameters in erosion rates and net momentum productions}

{\bf Physical parameters and dynamical variables involved in erosion rates:} Multi-phase mass flow and the associated erosion phenomena are characterized by distinct physical and mechanical parameters and flow dynamical variables. The erosion rate models (\ref{Eqn_11}) contain such parameters and variables from the flowing mixture and the  erodible basal substrate, and also the induced quantities such as the proportionality factor in the shear velocity, and erosion drifts which are expressed in terms of the stated parameters and variables. The physical and mechanical parameters are the densities, friction coefficients and Chezy-type friction coefficients. The dynamical variables are the volume fractions and the flow depth. Additional parameters are the local slope angle, and the typical flow depth ($H$), and depending on the choice of the basal shear stress, the effective pore pressure ratio may also inter the list of parameters. All these parameters are either measurable, or can be obtained from the literature, or are obtained from the mechanical closures derived above. The flow dynamical quantities are obtained directly from the model simulations. Full set of dynamical model is presented at Section 7. So, the mechanical erosion rate models in (\ref{Eqn_11}) are well defined and well constrained. 
\\[3mm]
In general, there are 
 six quantities that need to be closed in (\ref{Eqn_7}). These are, 
$\nu;
\lambda^m_l, \lambda^b$; $\lambda_{f_l}^m$; $\lambda_{ff}^b$ 
and $\lambda_{sf}^b$, 
which are the quantities and parameters
related to the shear velocity and the erosion drifts.
These quantities will be needed only if the solid and fluid erosion rates are derived explicitly for the solid and fluid phases. 
However, if we use 
(\ref{Eqn_4_4}) instead of (\ref{Eqn_3N}) then, we need to parameterize $\Upsilon^b$ instead of 
 $\lambda_{ff}^b$ and $\lambda_{sf}^b$. 
As shown in Section 3.1.3, except for $\Upsilon^b$, all these quantities can be closed or parameterized relatively easily. However, note that the legitimate mechanical closure for $\Upsilon^b$ is still an open question in mass flows (Iverson, 2012; Ouyang et al., 2015). So, the real multi-phase interactive shear structure (\ref{Eqn_3N}) is mechanically superior and dynamically flexible over the effectively single-phase reduced structure in (\ref{Eqn_4_4}) for the basal shear resistance.
Also, $\alpha_{_{BJ}}^m$, $\alpha_{_{BJ}}^b$; $C^m_f$, $C^b_f$ and $H$ are other physical parameters in $E$. However, $\alpha_{_{BJ}}^m$, $\alpha_{_{BJ}}^b$ can be modelled as explained in (\ref{Eqn_4NN}). Similarly, the values of $C^m_f$, $C^b_f$ can be obtained from the literature (Fraccarollo and Capart, 2002; Pudasaini and Fischer, 2020a). Moreover, the value of $H$ can be selected as a typical flow depth for a given event. 
So, the values of the parameters $\nu$; $C_f^m, C_f^b$; $H$, $\alpha_{_{BJ}}^m$, $\alpha_{_{BJ}}^b$, and $\mathcal K^m, \mathcal K^b$
can either be found in the literature or be closed as discussed above. 
Only $\mathcal K^m$ and $\mathcal K^b$ are new modelling parameters in the present model development. Otherwise, all the parameters in the erosion rates $\left (E_s, E_f\right )$ and the momentum productions $(\mathcal M)$ are already in Pudasaini and Fischer (2020a). However, the new models are mechanical much stronger, consistent and comprehensive than those in existing models. 
\\[3mm]
In total, there are five erosion drifts in the momentum productions (\ref{Eqn_14u})-(\ref{Eqn_15v}) including those in $ E$, namely, $\lambda^b, 
\lambda_{l}^m, \lambda_{f_{l}}^m,
\lambda_{sf}^b, \lambda_{ff}^b$.
In application, we only need 
$\lambda^b, 
\lambda_{sf}^b, \lambda_{ff}^b$ as the variation of the flow velocities through the depth can be neglected.
However, if (\ref{Eqn_6N}) and (\ref{Eqn_4_4}) are used instead of (\ref{Eqn_3N}) then, we only need $\lambda^b$
and $\Upsilon^b$, further reducing the erosion drifts by two, but as explained before, without clear mechanical closure for $\Upsilon^b$. If we use the momentum productions (\ref{Eqn_12u})-(\ref{Eqn_13v}), we effectively have five erosion drifts: $\lambda^b, \lambda_{sf}^b, \lambda_{ff}^b; \lambda_{fs}^b, \lambda_{ss}^b$, where $\lambda_{fs}^b, \lambda_{ss}^b$ emerge due to the composite erosion velocities. As explained earlier, in any situation, all these erosion drifts are mechanical closed.
\\[3mm]
We mention that one cannot expect well-defined mechanical models without the involvement of a number of intrinsic physical parameters characterizing the dynamics and the complexity of natural events. However, the parameter fit approach based on the empirical models cannot be anticipated to meaningfully represent the complex natural phenomenon of erosive mass transport. With the mechanically-explained models, such as the ones developed here, one understands a wide spectrum of natural processes taking place in a deterministic way, which is far beyond the reach of empirical models.  

\subsection{Recovering the existing model}

 In the limits, we recover the solid and fluid erosion rates
in Pudasaini and Fischer (2020a) by respectively neglecting the fluid contribution and
the solid contribution from the unified erosion rate model developed in (\ref{Eqn_7}).
However, the erosion rates in Pudasaini and Fischer (2020a) could not be directly
 generalized to obtain the unified erosion rate (\ref{Eqn_7}). It was made possible here 
  by following the process of considering total stresses, mixture densities, shear velocity
  and erosion drift equations for the mixtures and other
 aspects as discussed in Section 2 - Section 5. 
 
 \subsection{Extension to multi-phase erosion rates}

From the derivations and structures of the models developed above, it is straightforward to observe that,
our method can be directly extended to derive the erosion rates for
multi-phase mass flows, consisting of any number of constituent solid
particles and viscous fluid phases in the flow material and the bed
substrate. A particular example is the three-phase mass flows consisting of the coarse solid, fine solid and the fluid phases (Pudasaini and Mergili, 2019).
 All the properties of the multi-phase erosion rate (mass production) models explained above also apply to the multi-phase momentum productions.
 
\section{Full dynamical model with unified erosion rates and\\ net momentum productions}
   
In two-phase debris mixtures, phases are characterized by different material properties. The fluid phase is characterized by its material density $\rho_{f}$, viscosity $\eta_{f}$  and isotropic stress distribution; whereas the solid phase is characterized by its material density $\rho_{s}$, particle diameter $d$, the internal friction angle $\phi$, the basal friction angle $\delta$, an anisotropic stress distribution, and the lateral earth pressure coefficient $K$. The subscripts $s$ and $f$ represent the solid and the fluid phases respectively, with the depth-averaged velocity components for fluid  $\textbf{u}_{f}$ = ($u_{f}$, $v_{f}$) and for solid $\textbf{u}_{s}$ = ($u_{s}$, $v_{s}$) in the down-slope $(x)$ and the cross-slope $(y)$ directions. The total flow depth is denoted by $h$, and the solid volume fraction $\alpha_s$ (similarly the fluid volume fraction $\alpha_f = 1- \alpha_s$) are functions of space and time. Note that, except for the mass and momentum production rates, all other variables and parameters in the mass and momentum balance equations and other related terms, including forces, are for the landslide mixture. However, in what follows, for notational convenience, the superscript $^m$ indicating the mixture quantities have been omitted. 
\\[3mm]
The solid and fluid mass balance equations for the landslide (Pudasaini, 2012) including the mass productions (erosion rates) together with the evolution equation for the basal morphology are given by
\begin{eqnarray}
\begin{array}{lll}
\D{\Pd{}{t}{\lb \alpha_s h\rb} + \frac{\partial}{\partial x}{\lb \alpha_s h u_s\rb}
                 + \frac{\partial}{\partial y}{\lb \alpha_s h v_s\rb}=E_s},
\\[5mm]
\D{\Pd{}{t}{\lb \alpha_f h\rb} + \frac{\partial}{\partial x}{\lb \alpha_f h u_f\rb}
                 + \frac{\partial}{\partial y}{\lb \alpha_f h v_f\rb}=E_f,}
\\[5mm]
\D{\Pd{b}{t}= -E; \,\,\,\,\,\,\, E = E_s + E_f,}
\end{array}    
\label{Model_Final_Mass}
\end{eqnarray}
where $b = b(x,y; t)$ is the basal topography that evolves in space $(x, y)$ and time $(t)$, and $E_s$, $E_f$ are the solid and the fluid erosion-rates, and $E$ is the total erosion-rate as given by (\ref{Eqn_11}) and (\ref{Eqn_7}), respectively. This model can be used for partially or, fully saturated erodible basal substrate, or the substrate that is not erodible ($E = 0$). When the basal substrate is erodible, the solid fraction of $E$, i.e., $E_s$, enters into the solid mass balance as the solid mass production. So does the fluid fraction of $E$, i.e., $E_f$, that enters into the fluid mass balance as the fluid mass production. 
\\[3mm]
 Similarly, momentum conservation equations for the solid and fluid phases, in the down-slope ($x$) and cross-slope ($y$) directions, respectively, are: 
\begin{eqnarray}
\begin{array}{lll}
\resizebox{.935\hsize}{!}{$\D{\Pd{}{t}\biggl [ \alpha_s h \lb u_s \!-\! \gamma \mathcal C\lb u_f\! -\!u_s \rb \rb \biggr ]
  \!+\!\Pd{}{x}\biggl [ \alpha_s h \lb u_s^2 \!-\! \gamma \mathcal C\lb u_f^2 \!-\!u_s^2 \rb\!+\! \beta_{x_s} \frac{h}{2}\rb \biggr ]
  \!+\!\Pd{}{y}\biggl[ \alpha_s h \lb u_sv_s \!-\! \gamma \mathcal C\lb u_fv_f \!-\!u_sv_s \rb \rb \biggr ]}
  \D{=  h\mathcal S_{x_s} \!+\! 2 \mathcal M_{x_s}}$},\\[5mm]
\resizebox{.935\hsize}{!}{$\D{\Pd{}{t}\biggl [ \alpha_s h \lb v_s \!-\! \gamma \mathcal C\lb v_f \!-\!v_s \rb \rb \biggr ]
  \!+\!\Pd{}{x}\biggl [ \alpha_s h \lb u_sv_s \!-\! \gamma \mathcal C\lb u_fv_f \!-\!u_sv_s \rb\rb \biggr ]
  \!+\!\Pd{}{y}\left[ \alpha_s h \lb v_s^2 \!-\! \gamma \mathcal C\lb v_f^2 \!-\!v_s^2\rb\!+\! \beta_{y_s} \frac{h}{2}  \rb \right ]}
\D{=  h\mathcal S_{y_s}\!+\! 2 \mathcal M_{y_s}}$},\\[5mm]
\resizebox{.935\hsize}{!}{$ \D{\Pd{}{t}\left [ \alpha_f h \lb u_f \!+\! \frac{\alpha_s }{\alpha_f}\mathcal C\lb u_f \!-\!u_s \rb \rb \right ]
  \!+\!\Pd{}{x}\left [ \alpha_f h \lb u_f^2 \!+\! \frac{\alpha_s }{\alpha_f}\mathcal C\lb u_f^2 \!-\!u_s^2 \rb  \!+\! \beta_{x_f} \frac{h}{2}\rb \right ]
  \!+\!\Pd{}{y}\left[ \alpha_f h \lb u_fv_f  \!+\! \frac{\alpha}{\alpha_f}\mathcal C\lb u_fv_f \!-\!u_sv_s \rb \rb \right ]
=  h\mathcal S_{x_f}\!+\! 2 \mathcal M_{x_f}}$},\\[5mm] 
\resizebox{.935\hsize}{!}{$ \D{\Pd{}{t}\left [ \alpha_f h \lb v_f \!+\! \frac{\alpha_s }{\alpha_f}\mathcal C\lb v_f \!-\!v_s \rb \rb \right ]
 \! +\!\Pd{}{x}\left [ \alpha_f h \lb u_fv_f \!+\! \frac{\alpha_s }{\alpha_f}\mathcal C\lb u_fv_f \!-\!u_sv_s \rb\rb \right ]
  \!+\!\Pd{}{y}\left[ \alpha_f h \lb v_f^2  \!+\! \frac{\alpha_s}{\alpha_f}\mathcal C\lb v_f^2 \!-\!v_s^2 \rb \!+\!  \beta_{y_f} \frac{h}{2}\rb \right ]
=  h\mathcal S_{y_f}\!+\! 2 \mathcal M_{y_f}}$},
\end{array}    
\label{Model_Final}
\end{eqnarray}
where $\mathcal S$ are the source terms (discussed below), including all the frictional, collisional and the viscous stress generating mechanisms mentioned above, and the momentum productions $\mathcal M$ are given by (\ref{Eqn_14u})-(\ref{Eqn_15v}) or (\ref{Eqn_12u})-(\ref{Eqn_13v}), respectively.
\\[3mm]
These solid and fluid momentum equations are rigorously derived (Pudasaini, 2012), and include the solid and fluid momentum production terms, as modelled in Pudasaini and Krautblatter (2021), second terms on the right hand sides. 
Following Pudasaini and Krautblatter (2021), the momentum balance equations (\ref{Model_Final}) correctly include the erosion-induced change in inertia and the momentum production of the system via the terms $2 \mathcal M$, which are the net momentum productions. Importantly, our present approach makes a complete description of the full dynamical model equations for multi-phase erosive landslide in conservative form by considering all the aspects associated with the erosion-induced reduced friction (the momentum production) and the correct handling of the inertia of the system.
One of the important aspects in these momentum production terms are, that the velocities of the solid and fluid particles at the bottom that have just been eroded, $u^b, v^b$, or $u_s^b, v_s^b; u_f^b, v_f^b$, that appear in the net momentum productions (\ref{Eqn_14u})-(\ref{Eqn_15v}) or (\ref{Eqn_12u})-(\ref{Eqn_13v}) are different than the depth-averaged (mean) velocities, $u, v$, or $u_s, v_s; u_f, v_f$, that appear in the inertial (or, the convective) part, and also the source terms, of the mass and momentum equations. 
\\[3mm]
In (\ref{Model_Final}), the source terms (written in dimensional form, Pudasaini and Mergili, 2019) are as follows:
\begin{eqnarray}
\begin{array}{lll}
\hspace{-17mm}\D{\mathcal S_{x_s} = \alpha_s\left [g^x - \frac{u_s}{|{\bf u}_s|}C\tan\delta g^z(1-\gamma) 
- (1-C)\lambda^2 d^2 \left(\frac{u_s}{h}\right)^2\frac{1}{h}
- g^z(1-\gamma)\Pd{b}{x}\right ] 
- \gamma\alpha_s g^z\!\left [ \Pd{h}{x} + \Pd{b}{x}\right ]}\\[5mm]
\hspace{-5mm} +\, \D{C_{DG} \lb u_f - u_s \rb{ |{\bf u}_f - {\bf u}_s|}^{\jmath-1} 
-C_{DV}^{x_s} u_s|{\bf u}_s| \alpha_s},
\end{array}   
\label{Model_Final_ss}
\end{eqnarray}
\begin{eqnarray}
\begin{array}{lll}
\hspace{-17mm}\D{\mathcal S_{y_s} = \alpha_s\left [ g^y - \frac{v_s}{|{\bf u}_s|}C\tan\delta g^z(1-\gamma) 
- (1-C)\lambda^2 d^2 \left(\frac{v_s}{h}\right)^2\frac{1}{h}
- g^z(1-\gamma)\Pd{b}{y}\right ] 
- \gamma \alpha_s g^z\!\left [ \Pd{h}{y} + \Pd{b}{y}\right ]}
\\[5mm]
\hspace{-5mm} +\, \D{C_{DG} \lb v_f - v_s \rb{ |{\bf u}_f - {\bf u}_s|}^{\jmath-1}
-C_{DV}^{y_s} v_s|{\bf u}_s|\alpha_s},
\end{array}   
\label{Model_Final_s}
\end{eqnarray}
\vspace{-3mm}
\begin{eqnarray}
\begin{array}{lll}
\D{\mathcal S_{x_f} = \alpha_f\biggl [g^x - \biggl [-\frac{1}{2}p_{b_f}\frac{h}{\alpha_f}\Pd{\alpha_f}{x} +  p_{b_f}\Pd{b}{x}
 -\left \{ 
 2\frac{\partial}{\partial x}\left ( \nu_f\frac{\partial u_f}{\partial x}\right )
 +\frac{\partial}{\partial y}\left ( \nu_f\frac{\partial v_f}{\partial x}\right )
 +\frac{\partial}{\partial y}\left ( \nu_f\frac{\partial u_f}{\partial y}\right )
 - \nu_f\frac{\chi u_f}{h^2} \right \}} \\[5mm]
 +\D{
 \frac{\mathcal A}{\alpha_f}\left \{ 
 2\frac{\partial}{\partial x}\left ( \nu_f \frac{\partial \alpha_s}{\partial x}\left( u_f-u_s\right )\right)
 +\frac{\partial}{\partial y}\left ( \nu_f 
 \left(\frac{\partial \alpha_s}{\partial x}\left( v_f-v_s\right )
 +\frac{\partial \alpha_s}{\partial y}\left( u_f-u_s\right )
 \right)\right)
 \right\}
 -\frac{\mathcal A}{\alpha_f}\frac{\xi\alpha_s\nu_f}{h^2}\left ( u_f -u_s\right)
 \biggr]\biggl ]}\\[5mm] 
-\D{\frac{1}{\gamma}C_{DG}\lb u_f - u_s \rb{ |{\bf u}_f - {\bf u}_s|}^{\jmath-1}
 -C_{DV}^{x_f} u_f|{\bf u}_f| \alpha_f},
\end{array}    
\label{Model_Final_fx}
\end{eqnarray}
\vspace{-3mm}
\begin{eqnarray}
\begin{array}{lll}
\D{\mathcal S_{y_f} = 
\alpha_f\biggl [g^y - \biggl [-\frac{1}{2}p_{b_f}\frac{h}{\alpha_f}\Pd{\alpha_f}{y} +  p_{b_f}\Pd{b}{y}
 -\left \{ 
 2\frac{\partial}{\partial y}\left ( \nu_f\frac{\partial v_f}{\partial y}\right )
 +\frac{\partial}{\partial x}\left ( \nu_f\frac{\partial u_f}{\partial y}\right )
 +\frac{\partial}{\partial x}\left ( \nu_f\frac{\partial v_f}{\partial x}\right )
 - \nu_f\frac{\chi v_f}{h^2} \right \}} \\[5mm]
 +\D{
 \frac{\mathcal A}{\alpha_f}\left \{ 
 2\frac{\partial}{\partial y}\left ( \nu_f \frac{\partial \alpha_s}{\partial y}\left( v_f-v_s\right )\right)
 +\frac{\partial}{\partial x}\left ( \nu_f 
 \left(\frac{\partial \alpha_s}{\partial y}\left( u_f-u_s\right )
 +\frac{\partial \alpha_s}{\partial x}\left( v_f-v_s\right )
 \right)\right)
 \right\}
 -\frac{\mathcal A}{\alpha_f}\frac{\xi\alpha_s\nu_f}{h^2}\left ( v_f -v_s\right)
 \biggr]\biggl ]}\\[5mm] 
-\D{\frac{1}{\gamma}C_{DG}\lb v_f - v_s \rb{ |{\bf u}_f - {\bf u}_s|}^{\jmath-1}
-C_{DV}^{y_f} v_f|{\bf u}_f| \alpha_f}.
\end{array}    
\label{Model_Final_fy}
\end{eqnarray}
The pressures and other parameters involved in the above model equations are as follows:
\begin{eqnarray}
\begin{array}{lll}
\D{
\beta_{x_s} = K_x g^z(1-\gamma), \,\,\,\, 
\beta_{y_s} = K_y g^z(1-\gamma),\,\,\,
\beta_{x_f} = \beta_{y_f} = g^z,\,\,\, p_{b_f} =  g^z, \,\,\, p_{b_s} = (1-\gamma)p_{b_f},}\\[5mm]
\D{C_{DG} = \frac{\alpha_s \alpha_f(1-\gamma)g}{\left [\mathcal U_T\{{\cal P}\mathcal F(Re_p) + (1-{\cal P})\mathcal G(Re_p)\} + {\mathcal S}_P\right ]^{\jmath}},\,\,\,\,
\mathcal F = \frac{\gamma}{180}\lb\frac{\alpha_f}{\alpha_s} \rb^3 Re_p, \,\,\,\,  \mathcal G= \alpha_f^{M(Re_p) -1},}
\\[5mm]
\D{\gamma =\frac{\rho_f}{\rho_s},\, Re_p = \frac{\rho_f d~ \mathcal U_T}{\eta_f},\, \nu_f = \frac{\eta_f}{\rho_f},\,
\alpha_f = 1-\alpha_s,\, \mathcal A = \mathcal A(\alpha_f).}
\end{array}    
\label{Model_Final_parameters}
\end{eqnarray}
 Equations (\ref{Model_Final_Mass}) are the depth-averaged mass balances for solid and fluid phases respectively, and (\ref{Model_Final}) are the depth-averaged momentum balances for solid (first two equations) and fluid (other two equations) in the $x$- and $y$-directions, respectively. All equations and expressions are written in dimensional form.
\\[3mm]
In the above {equations (\ref{Model_Final_Mass})-(\ref{Model_Final}), $x$, $y$ and $z$ are the locally orthogonal coordinates in} the down-slope, cross-slope and flow normal directions, and $g^x$, $g^y$, $g^z$ are the respective components of gravitational acceleration. 
 $\mu =\tan\delta$ is the basal friction coefficient, $\lambda$ is the Bagnold linear concentration, $d$ is the particle diameter, and 
$C_{DG}$ is the generalized drag coefficient. Simple linear (laminar-type, at low velocity) or quadratic (turbulent-type, at high velocity) drag is associated with ${\jmath} = 1$ or $2$, respectively. $\mathcal{U}_{T}$ is the terminal velocity of a particle, $\mathcal{P}\in [0,1]$ is a parameter, {or a function (Pudasaini, 2020)} which combines the solid-like ($\mathcal{G}$) and fluid-like ($\mathcal{F}$) drag contributions to flow resistance, and $C \in [0, 1]$ linearly combines the frictional and the collisional stresses of the solid particles. $p_{b_{f}}$ and $p_{b_{s}}$ are the effective fluid and solid pressures. $\gamma$ is the density ratio, $\nu_f$ is the kinematic viscosity of fluid, $\mathcal{C}$ is the virtual mass coefficient (kinetic energy of the fluid phase induced by solid particles, {(Pudasaini, 2019)}), $M$ is a function of the particle Reynolds number ($R_{e_{p}}$), $\chi$ includes vertical shearing of fluid velocity, $\xi$ takes into account different distributions of $\alpha_s$, and $\mathcal{A}$ is the mobility of the fluid at the interface between the solid and fluid in the flow.
{$C_{DV}$ are the viscous drag coefficients (Pudasaini and Hutter, 2007), akin to Chezy-friction, that can also include the high intensity frontal ambient drag (Pudasaini and Fischer, 2020a)}. 
\\[3mm]
The physically based, fully analytical, and well-bounded two-phase virtual mass force coefficient $\mathcal C$ (Pudasaini, 2019) is given by:
$\mathcal C = \frac{\mathcal N_{vm}^0\left ( \l\, + \,\alpha_s^q\right) - 1}{\left(\alpha_f/\alpha_s\right) +\, \gamma}$,
where $\mathcal N_{vm}^0$ is the virtual mass number, $\l$ and and $q$ are some numerical parameters. This model covers any distribution of the dispersive phase (dilute to dense distribution of the solid particles). As justified in Pudasaini (2019), the physically relevant values of these parameters can be $\mathcal N_{vm}^0 = 10.0, \l = 0.12, q = 1$. This virtual mass force is general and evolves automatically as a function of solid volume fraction.
\\[3mm]
$\mathcal S_P = \left ( \frac{P}{\alpha_s} + \frac{1 - P}{\alpha_f}\right ){\mathcal K}$ is called the smoothing function, where ${\mathcal K}$ is determined by the corresponding mixture mass flux per unit mixture density, typically 10 ms$^{-1}$ (Pudasaini, 2020). $P = \alpha_s^n$ is a function of the solid volume fraction $\alpha_s$, where $n$ is a positive number, close to 1.
\\[3mm]
 The evolution of basal topography ${\partial b}/{\partial t}= -E$ in (\ref{Model_Final_Mass}) due to erosion and deposition is explicitly included in the model. With this, the basal change directly influences the source terms in (\ref{Model_Final_ss})-(\ref{Model_Final_fy}) by accounting for changes that are associated with the driving {and resisting} forces in the net force balance. This appears very important for geophysical mass flows which are mainly driven by gravity and slope changes, i.e., the respective components of gravitational accelerations, frictions, the basal and hydraulic pressure gradients, and the buoyancy induced terms.
\\[3mm]
In the derivation of the model in this section, extending Pudasaini and Fischer (2020a), we assumed that the solid obeys to the frictional Coulomb and the Bagnold collisional laws at the base of the flow, and also in the bed when dealing with the erosion rate. However, at very
large particle concentrations, the Coulomb rheology could be generalized by applying rate-dependent
particle stresses such as the phenomenological models based on the $\mu(I)$ rheology (dry confined flows
by Jop et al., 2005; for water-immersed grains by Cassar et al., 2005; confined submarine avalanches
by Doppler et al., 2007), or even more complex pressure- and rate-dependent Coulomb-viscoplastic
rheologies (Domnik et al., 2013; Pudasaini and Mergili, 2019). Moreover, fundamental approaches
based on kinetic theory, which shows that the particle stresses are rate-dependent, might produce appreciable
results for erosion associated with large particle concentration (Berzi and Fraccarollo, 2015). So, the
models presented here can further be extended and generalized to the situation that the stresses are rate-dependent that might better deal with the erosion phenomena for large particle concentration.
\\[3mm]
As discussed in Pudasaini and Fischer (2020a), in general, the above model can be applied to the fluid of any viscosity, where the effective viscosity increases
with the particle concentration (Takahashi, 2007; Pudasaini, 2012; Pudasaini and Mergili, 2019). We
have applied both the Coulomb and the Bagnold stresses for particle or solid-phase. However, we note that, for the laminar flows,
the particle stresses can scale with the fluid viscosity, as in viscous suspensions (Ness and Sun, 2015).
For more complex situation with turbulence, e.g., turbulence in the presence of particles, and the
contribution of the particle fluctuations to the fluid viscosity, we refer to Berzi and Fraccarollo (2015)
who
showed that turbulent viscosity is a decreasing function of the local
volume fraction of particles.
Structurally, the mass and momentum equations  (\ref{Model_Final_Mass}) and (\ref{Model_Final})  are the same as in Pudasaini and Fischer (2020a). However, there are fundamental differences, due to the applied erosion rate models in Pudasaini and Fischer (2020a) and here, and the new net momentum productions.

\section{Erosion-induced net momentum production and flow mobility}

\begin{figure}
\begin{center}
\includegraphics[width=9cm]{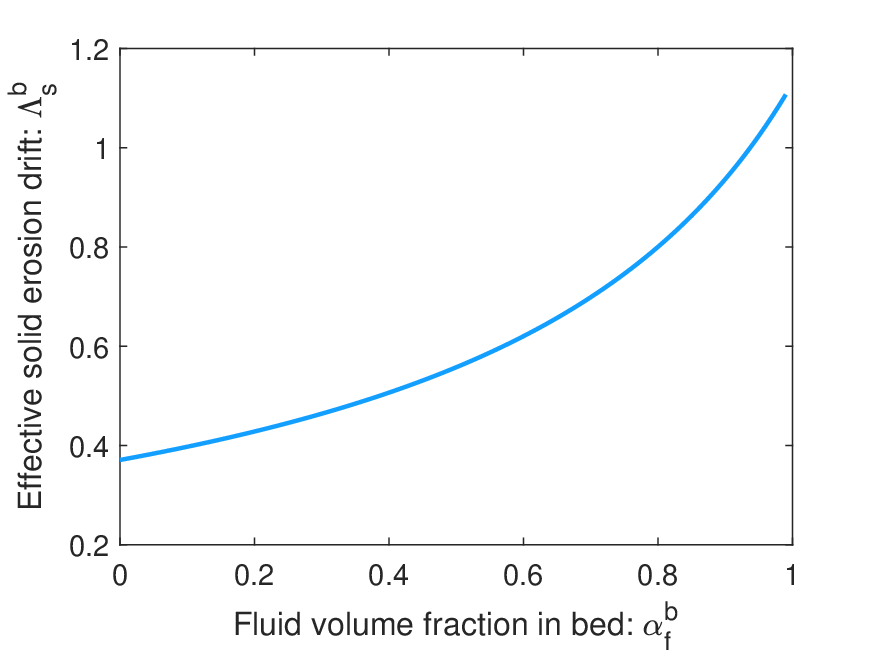}
\includegraphics[width=9cm]{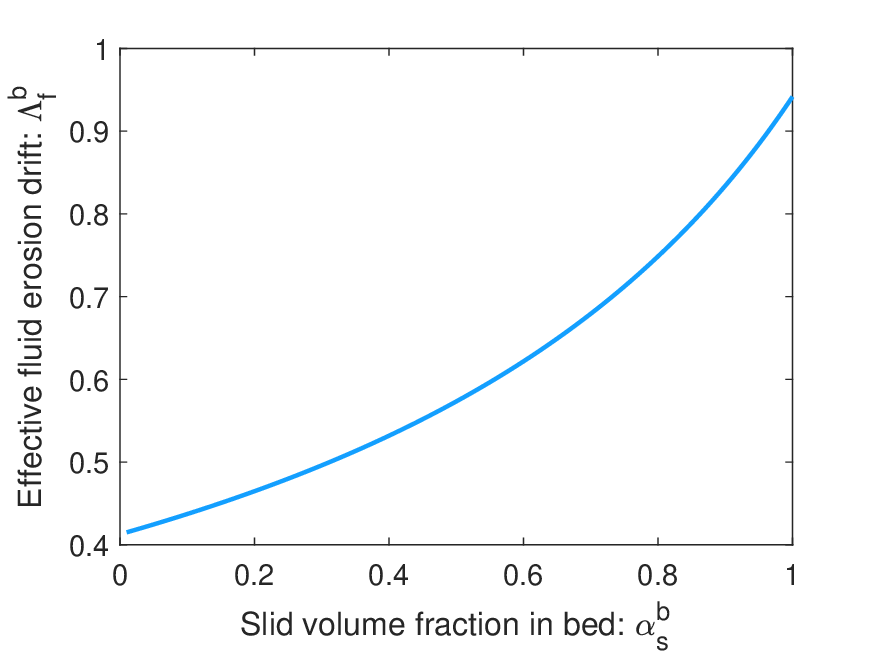}
  \end{center}
  \caption[]{The effective composite erosion drifts as given by (\ref{Eqn_ubs_alt}) and (\ref{Eqn_ubf_alt}) for the composite solid and fluid erosion velocities $u_s^b$ and $u_f^b$, respectively.}
  \label{Fig_7}
\end{figure}

Pudasaini and Krautblatter (2021) presented a mechanical condition for when, how and how much energy erosive landslides gain or lose. They pioneered a mechanical model for the energy budget of erosive landslides that controls enhanced or reduced mobility, 
and made a breakthrough in correctly determining the landslide mobility. Erosion velocity, which regulates the energy budget, determines the enhanced or reduced mobility. 
With their energy generator they offered the first-ever mechanical quantification of erosional energy and a precise description of mobility. 
They demonstrated that erosion and entrainment are different processes.
Landslides gain energy and enhance mobility if the erosion velocity $\left(u^b\right)$ exceeds the entrainment velocity $\left( u - u^b\right)$. 
Presented dynamical equations in  Pudasaini and Krautblatter (2021) correctly include erosion induced net momentum production. However, the Pudasaini and Krautblatter (2021) erosion-induced landslide mobility model is for an effectively single-phase bulk mixture, that we have extended here and applied in (\ref{Model_Final_Mass})-(\ref{Model_Final}) for a true two-phase (multi-phase) mass flows with unified mechanical model for erosion rates, and the rates of mass and momentum productions in (\ref{Eqn_11}) and in (\ref{Eqn_14u})-(\ref{Eqn_13v}).   
Yet, the mixture model presented here is more realistic, but also more comprehensive than in Pudasaini and Krautblatter (2021) due to the extended structure of erosion velocities, unified mechanical erosion rates, and the associated advanced net momentum productions in (\ref{Eqn_14u})-(\ref{Eqn_13v}), all based on the extensive description of the frictional, collisional and viscous stress generating mechanisms.
\\[3mm]
It is important to note that, in (\ref{Model_Final}) out of the erosion induced net momentum productions $2 \mathcal M$, one $\mathcal M$ emerges from the momentum production derived from the effectively reduced friction, while the other $\mathcal M$ originates from the correct understanding of the inertia of the entrained mass. Pudasaini and Krautblatter (2021) invented these crucial aspects. Mechanically and dynamically, this makes a huge difference, and thus, is a great advancement in simulating landslide with erosion. 
However, as explained above, the structure and scope of $\mathcal M$ here is much more extensive than that in Pudasaini and Krautblatter (2021).
\\[3mm]
Pudasaini and Krautblatter (2021) proved that, if the erosion velocity is greater than one-half of the flow velocity the mobility is enhanced. For the momentum productions in (\ref{Eqn_14u})-(\ref{Eqn_15v}), this is $u^b > u^m/2$, or equivalently $\lambda^b > 1/2$. In other words, the landslide gains energy to enhance its mobility if the eroded material is easily entrainable with the velocity lower than the erosion velocity. Otherwise, the mobility of the mass flow will be reduced even for the erosive events. For the momentum productions in (\ref{Eqn_12u})-(\ref{Eqn_13v}), in principle, these conditions are valid in terms of the erosion velocities in connection with the flow velocities
and the erosion drifts. However, the erosion velocities and the erosion drifts need to be restructured.
The erosion velocities $u_s^b$ and $u_f^b$ in (\ref{Eqn_12u})-(\ref{Eqn_13u}) can be re-written in convenient forms as:
\begin{equation}
u_s^b  
= \left[\alpha_s^m \lambda_{ss}^b u_s^m + \alpha_f^m \lambda_{fs}^b u_f^m\right]
= \alpha_s^m\left[ \lambda_{ss}^b\left\{ 1 + \frac{\alpha_f^m \lambda_{fs}^b u_f^m}{\alpha_s^m \lambda_{ss}^b u_s^m}\right\}\right]u_s^m
= \alpha_s^m \hat{\lambda}_{s}^b u_s^m
= \Lambda_{s}^b u_s^m,
\label{Eqn_ubs_alt}
\end{equation}
\begin{equation}
u_f^b  
= \left[\alpha_f^m \lambda_{ff}^b u_f^m + \alpha_s^m \lambda_{sf}^b u_s^m\right]
= \alpha_f^m\left[ \lambda_{ff}^b\left\{ 1 + \frac{\alpha_s^m \lambda_{sf}^b u_s^m}{\alpha_f^m \lambda_{ff}^b u_f^m}\right\}\right]u_f^m
= \alpha_f^m \hat{\lambda}_{f}^b u_f^m
= \Lambda_{f}^b u_f^m.
\label{Eqn_ubf_alt}
\end{equation}
In these representations, the solid and fluid mobilities will be enhanced if
$u_s^b > u_s^m/2, u_f^b > u_f^m/2$, or
$\Lambda_{s}^b > 1/2$ and $\Lambda_{f}^b > 1/2$, which can be considered component-wise or together. Otherwise, the mobility will be reduced if $\Lambda_{s}^b < 1/2$ and $\Lambda_{f}^b < 1/2$, and remains unchanged if $\Lambda_{s}^b = 1/2$ and $\Lambda_{f}^b = 1/2$ even for erosive mass flows.
This way, we have presented extended conditions for the erosion-induced net momentum production and flow mobility as required for the mixture mass flows.
\\[3mm]
 The effective composite solid and fluid erosion drifts appearing in connection with the composite erosion velocities for solid and fluid are presented in Fig. \ref{Fig_7}. These composite erosion drifts characterize the solid and fluid erosion velocity with the single phase-type erosion drifts. However, depending on the chosen solid and fluid volume fractions in the landslide and other physical quantities across the erosion interface, these erosion drifts are structurally and mechanically different, and may also be higher than the other phase- and cross-phase erosion drifts, which are bounded from above by unity. This is surprising.

\section{Discussions}

{\bf I. Complex phase- and cross-phase interactions:}
 The astonishing fact that emerged here while dealing with the erosion velocities of the mobilized bed materials is that there are four fundamentally different types of phase-phase and cross-phase interactions across the erosion interface between the solid and fluid materials in the landslide and the bed: 
 the direct solid-solid and fluid-fluid interactions, and the cross solid-fluid and fluid-solid interactions.
 Because, the 
existing erosion models only consider the direct 
phase-interactions, 
 cross-phase interactions are not recognized yet, limiting their applicabilities,
as those models are incomplete and partly inconsistent. 
The fluid-fluid interactions are not described appropriately as they must follow some special mechanical principles which are freshly realized here. Similarly, the true solid-fluid and fluid-solid interactions are ignored previously although these interactions are substantial and appear to be quite complex. 
\\[3mm]
{\bf II. Interactive phase- and cross-phase shear stresses:}
 When both the landslide and erodible bed are composed of two-phase materials of different physical properties, the interface shear stresses between them become very complicated. 
In previous erosion models, the fluid-solid and solid-fluid shear stresses do not exist as the cross-shear stresses, but rather were directly used as fluid-fluid interactions, which are physically inconsistent. Moreover, the fluid-fluid shear stresses do not exist at all. Here, we made a fundamental advancement in modelling multi-phase erosive mass transport with entirely innovative and appropriate mechanical 
shear stress models for all the solid-fluid, fluid-solid and fluid-fluid interactions
across the erosion-interface.
We introduced mechanically important inter-phase solid-fluid and fluid-solid interactions. We revealed that the shear resistance by the fluid at the bed against the applied shear from the solid from the flow is rather associated with the solid velocity in the landslide but not with the fluid velocity at the bed. This is a novel understanding. 
\\[3mm]
The solid-solid interactions are modelled by combining the Coulomb-type frictional law with the Bagnold-type collisional law.
However, we presented elegant cross-couplings that are unique and legitimate. 
The solid-fluid interactions (the fluid-type bed shear resistance against the solid-type shear stress from the landslide) satisfy the Chezy-type frictional rheology, but with some crucial amendments.
Such special circumstance is invented here for erosive mass transports.
Astonishing fact we discovered is the description of the fluid-fluid interactions at the interface. 
Previously, fluid-fluid interactions were modelled with the classical Chezy-type friction which is mechanically inconsistent. However, these fluids are contained inside the matrices of the solid particles in the landslide and in the bed. As the fluid-fluid interactions are not yet known for the erosive landslide, following a boundary layer approach, we constructed new physically-based models. 
We described fluid shear stresses across the interface by assuming both the landslide and bed as porous medium. 
These shear stresses are distinguished by their viscosities and velocities, however, with permeabilities on opposite sides of the erosion-interface. 
Fluid resistances from the bed against the solid and fluid shear stress from the landslide are incomparably different. This is an important novel development. Physical mechanisms of all the interfacial shear stresses are justified.
Depending on the composition of the flowing landslide and the bed mixture and the significance of the relevant shear stresses, as recognized here, amazingly there are nine different shear stress jumps across the erosion interface. 
These aspects clearly manifest the physical novelty and significance of our modelling approach 
for the interactive shear stresses between different materials across the landslide-bed interface,
presented here for the first time for the complex multi-phase erosive mass transports.
This indicates the complexity associated with the two-phase erosive landslide, the novelty and essence of our approach. 
\\[3mm]
{\bf III. Composite erosion velocities:}
 Surprisingly, extended erosion velocities for both the solid particles and fluid molecules take mechanically extensive and complex forms. We revealed the real mechanical situations that the solid and fluid at the bed are mobilized both by the solid and fluid in the flow. 
With crucial novel realizations, we constructed comprehensive solid and fluid erosion velocities by considering all the
interactions and
mobilization components induced by both the solid and fluid from the flow to the solid and fluid in the erodible bed. 
There are composite contributions to the erosion velocities from the solid and fluid fractions from the flow and their velocities, and  drifts. 
The structures of erosion velocities clearly indicate that the process of erosion is dominated by the erosion velocities rather than the flow velocities themselves, which are the key findings.
These erosion velocities have huge implications in correctly describing the erosion-induced net momentum productions, because, the net momentum production entirely controls the dynamics, mobility, impact energy and the deposition morphology of the mass transport.   
\\[3mm]
{\bf IV. Unified, extensive and consistent erosion rates:}
Previously, the true solid-fluid and fluid-solid interactions were ignored and not described mechanically appropriately,
fiercely limiting the applicability of the existing erosion models.
Natural erosion rates that correspond to the events could not be obtained from the existing solid and fluid erosion rates. 
We solved these problems by presenting novel, mechanical shear stress models for the solid-fluid and fluid-fluid interactions.
 As erosion rates play a central role in erosive mass transports,
we focused on constructing a novel, unified and physically consistent extensive mechanical erosion rates for multi-phase mass flows.
Erosion rate is determined with the jump in the shear stresses and the jump in the momentum fluxes across the landslide-bed interface. 
The new unified erosion rate models include all mechanically and dynamically important interactions between the solid and fluid phases across the erosion-interface. 
The total basal erosion rate is extensive, compact, and is mechanically fully described. 
 The total erosion rate is the exact sum of the solid and fluid erosion rates, automatically satisfying the natural criterion as required by practitioners. 
 These erosion rates consistently take the solid and fluid fractions from the bed and customarily supply them to the flow. 
 This is crucial.
 We recover the solid and fluid erosion rates
in previous models. Our method can be directly extended to erosive
multi-phase mass flows consisting of any number of solid
particles and viscous fluid phases in the landslide and the bed
substrate. 
Importantly, we invented a novel erosive-shear-velocity primarily induced by the erosion rate, which vanishes for non-erosive flows.
\\[3mm]
{\bf V. Super-erosion-drift, phase- and cross-phase drifts:}
 As essential quantities, erosion drifts provide crucial information about the erosion velocities which play central role in explaining erosion rates and net momentum productions. In turn, net momentum productions control the mobility of erosive mass transports. We constructed different erosion drift equations providing mechanical closures for all the erosion drifts. 
We presented a compact and general super-erosion-drift-equation.
In the limit, it reduces to the solid-solid and fluid-fluid erosion drifts.
With elegant procedures, we also constructed closures for the solid-fluid and fluid-solid cross-erosion-drifts. 
All drifts are known mechanically.
We proved that 
 as the super-erosion-drift contains all necessary information, essentially all the phase- and cross-phase drifts can be directly extracted from the super-erosion-drift. The cross-drift are symmetrical about the solid-fluid and fluid-solid cross-phase interactions. These properties signify the strength of the super-erosion-drift and consistency of all the drift relations, and the equivalence between the reduced frictional forces and momentum productions.
\\[3mm]
{\bf VI. Complete net momentum productions and flow mobility:}
 The  very crucial aspect considered here for the first time is, as momentum productions play decisive role in the dynamics, mobility, destructive power and deposition morphology, we must mechanically correctly describe momentum productions with respect to the complex and compact erosion velocities and erosion rates. 
 We constructed the erosion-induced produced solid and fluid momenta in terms of the total erosion velocities, or in terms of the solid and fluid erosion velocities. 
  These momentum productions explicitly depend on several aspects of the flow: 
 erosion drifts, 
volume fractions of solid and fluid in the landslide
and the erodible bed, 
 solid and fluid velocities in the flow, and 
the total erosion rate of the system. 
 With the mechanical closures for the cross-erosion drifts, we revealed that it is relatively difficult for the fluid in the landslide to mobilize the grain in the bed, but it is relatively easy for the grain in the landslide to mobilize the fluid in the bed.
  However, our scrutiny shows that the erosion velocities 
are determined collectively by the involved erosion drifts, volume fractions of solid and fluid in the landslide, and their respective velocities. 
The newly constructed momentum productions reduce to the previously known solid only and fluid only momentum productions, the latter, however, are incomplete. This implies the wide spectrum and physically fully consistent 
 modelling of momentum productions in our approach associated with the solid and fluid erosion velocities. 
 This sheds light on the importance of the composite erosion velocities, unified erosion rates and the extensive net momentum productions for solid and fluid.
Existing erosion-induced landslide mobility model is only for an effectively single-phase bulk mixture, that we have extended here for multi-phase mass flows with unified mechanical model for the rates of mass and momentum productions.   
Erosion-induced net momentum productions here are much more extensive than that in the existing model, the former adequately describe the flow mobility of mixture mass flows.
\\[3mm]
{\bf VII. Erosion-matrix and process of erosion:}
 We invented the erosion-matrix characterizing the erosion mechanism of the landslide. 
 With the erosion matrix, we have presented the first systematic, compact and the complete description of the  mechanical process of erosion.
 As the erosion velocity governs the system, the process of erosion is jointly determined by the erosion-matrix, volume fractions of solid and fluid in the flow, and the flow velocities. 
 For the vanishing off-diagonal elements of the erosion-matrix, the system degenerates to the previously known simple erosion velocities, but without the cross-phase interactions, which is incomplete. 

\section{Summary}

 There are four major outcomes of this contribution in relation to erosive multi-phase mass flows. 
 First, we physically correctly established the jumps in shear stresses and momentum fluxes across the erosion-interface between the landslide and the bed substrate, and with these, we constructed unified, comprehensive and consistent mechanical erosion rates for the solid and fluid phases. The general structure of these jumps demonstrate the richness, urgency and spectrum of applicability of the new unified multi-phase erosion model. 
 The constructed shear resistances from the bed against all the applied shear stresses from the landslide are consistent and appropriate that include the frictional, collisional and viscous stress generating mechanisms. 
 The proposed  multi-phase interactive shear structures are mechanically explained, which are physically superior, dynamically flexible and wider over the existing effectively single-phase shear structures as many of the interactions considered here could not be described by existing models. 
 In our approach, the sum of the solid and fluid erosion rates turns out be the total basal erosion rate. The new erosion rate models are well defined and well constrained which can be applied to any flow situations and bed morphologies, irrespective of the number of components in the flow and the bed, and their interactions. Such a broad erosion modelling is presented here for the first time with our unified modelling approach. 
 Second, we constructed extensive and complete net momentum productions for both the solid and fluid phases 
 for which we completely and physically correctly described the essentially complex, composite erosion velocities of the mobilized particles and fluid from the basal substrate and the unified erosion rates. 
 Mass and momentum productions include all the interactions between the solids and fluids in the landslide and the bed substrate, which are consistent and mechanically much stronger and wider than the existing models. 
 Third, a general frame of the mass and momentum balance equations has been presented. 
 We pioneered the stress correction factor, erosive-shear-velocity, super-erosion-drift, and the erosion-matrix. These greatly enhance our understanding by inherently characterizing the complex erosion processes in multi-phase mass flows. 
  Fourth, we developed a realistic and comprehensive multi-phase mechanical erosion model by embedding all the frictional, collisional and viscous stresses for the solid and fluid phases, the novel and unified mechanical erosion rates, extended erosion velocities and the advanced net momentum productions into the mass and momentum balance equations. This removes the great hurdle in existing erosion modeling, and opens a wide spectrum of possibilities for real applications. Our approach makes a complete description of the full multi-phase erosive landslide in conservative form by considering all the aspects associated with the erosion-induced momentum productions and the correct handling of the inertia of the system via net momentum productions. 
 The mechanically-explained models developed here cover a vast range of natural processes in a deterministic way, which is far beyond the reach of empirical models. This paves the way for the legitimate applications of the developed erosion model for complex multi-phase mass flows. So, the professionals and engineers may find the model intuitive and useful in solving applied, technical, engineering and geomorphological problems associated with erosive mass flow events.

\end{document}